\documentclass[]{elsarticle}

\usepackage{amsmath}
\usepackage[colorlinks]{hyperref}

\journal{Icarus}

\biboptions{authoryear}

\usepackage[right=3cm,left=3cm]{geometry}
\begin{document}

\begin{frontmatter}

\title{Role of the global water ocean on the evolution of Titan's primitive atmosphere}

\author[lpg,chicago]{Nadejda Marounina\corref{mycorrespondingauthor}}
\cortext[mycorrespondingauthor]{Corresponding author}
\ead{nmarounina@gmail.com}

\author[lpg]{Olivier Grasset}
\author[lpg]{Gabriel Tobie}
\author[lpg]{Sabrina Carpy}

\address[lpg]{LPG-Nantes, 2 rue de la Houssiniere, BP 92208 44322 Nantes Cedex 3, France }
\address[chicago]{Department of Astronomy and Astrophysics, 5640 S Ellis Ave., Chicago, IL 60637, USA}

\begin{abstract}
During the accretion of Titan, impact heating may have been sufficient to allow the global melting of water ice and the release of volatile compounds, mainly constituted of CO$_2$, CH$_4$ and NH$_3$. The duration and efficiency of exchange between the primitive massive atmosphere and the  global impact-induced water ocean likely play a key role in the chemical evolution of the early Titan's atmosphere. To investigate the atmospheric composition of early Titan for a wide range of global (atmosphere + ocean) composition in volatils, we first developed a gas-liquid equilibrium model of the NH$_3$-CO$_2$-H$_2$O system, where the non-ideal behavior of both gas and liquid phases, and the speciation of volatiles dissolved in the aqueous phase are taken into account. We show that the relative abundance of CO$_2$ and NH$_3$ determine the composition of Titan's atmosphere. For CO$_2$/NH$_3 \leq 1$ , CO$_2$ is massively dissolved in the ocean. On the contrary, for CO$_2$/NH$_3 > 1$, CO$_2$ is the main constituent of Titan's primitive atmosphere while the NH$_3$ atmospheric content is dramatically decreased. We then investigate the conditions for the formation of CH$_4$-rich clathrates hydrates at Titan's surface that could be the main reservoir of methane for the present-day atmosphere. In absence of reliable experimental data in the CH$_4$-CO$_2$-NH$_3$-H$_2$O system, the dissolution of methane in water is included using a simplified Henry's law approach. We find that if the concentration of CH$_4$ in Titan's building block was higher than $\sim$0.1 mol.kg$^{-1}$ and CO$_2$/NH$_3 < 3$, a large fraction of methane may be stored in the primordial crust, which would form at temperature below $\sim$10$^{\circ}$C.
\end{abstract}


\end{frontmatter}


\section{Introduction}
 \label{sec_intro}
Titan is the only satellite in the Solar System to possess a massive atmosphere that is mainly composed of N$_2$ ($\sim$98\%) and CH$_4$ ($\sim$2\%) \citep[e.g.][]{Niemann2010, Griffith2013}. Thanks to the Cassini-Huygens mission, the atmosphere of Titan has now been extensively studied. Still, the origin of this atmosphere is highly debated. Ammonia has long been advocated as the primordial source of nitrogen on Titan \citep{Atreya1978, Prinn1981, Lunine1987}. The Huygens probe provided key constraints for the origin of nitrogen (and also carbon) in Titan's atmosphere \citep{Niemann2005, Niemann2010}, which favor a NH$_3$ into N$_2$ oxidation mechanism. A careful analysis of the rare gas abundances in Titan's atmosphere shows that it is possible for a small fraction of the present-day atmosphere to be residual gas from the primitive nebula \citep{Glein2017}. However, the low $^{36}$Ar/N$_2$ abundance measured by the Gas Chromatograph Mass Spectrometer (GCMS) onboard ($\sim 2.7 \times 10^{-5}$, which is $\sim 3\times 10^5$ times smaller that the solar value, \citet{Niemann2010}) indicates that the primordial blocks that formed Titan contained a very small fraction of $^{36}$Ar and N$_2$ and that nitrogen has been brought on Titan in the form of easily condensable N-bearing compounds, most likely NH$_3$ \citep{Atreya2010}. The $^{14}$N/$^{15}$N ratio measured in N$_2$ in Titan's atmosphere (167.7, \citet{Niemann2010}) appears also very close to the value recently inferred in NH$_2$ radicals in comets produced by photodissociation of NH$_3$ \citep{Rousselot2014, Shinnaka2014}, suggesting a similar origin.\\
Several mechanisms have been proposed to explain the conversion of NH$_3$ into N$_2$ and the formation of the N$_2$-rich Titan's atmosphere: photochemical conversion \citep{Atreya1978}, impact-induced conversion either in the atmosphere \citep{McKay1988, Ishimaru2011} or in a NH$_3$-enriched icy crust \citep{Sekine2011, Marounina2015}, and endogenic processes \citep{Glein2009, Tobie2012, Glein2015a}. An efficient conversion of NH$_3$ into N$_2$ in the primitive atmosphere requires a NH$_3$-saturated warm atmosphere \citep{Atreya1978, McKay1988}, conditions which may have been met at the end of Titan's accretion when a large fraction of ices incorporated in Titan's interior was molten \citep{Kuramoto1994, Monteux2014}.  \\
During Titan's accretion, impact heating may have been sufficient to allow the global melting of water ice leading to, the formation of a thick water ocean in contact with the atmosphere constituted by the release of volatile compounds, mainly CO$_2$, CH$_4$ and NH$_3$ \citep{Lunine1987, Tobie2012, Monteux2014}. As NH$_3$ is highly soluble in water \citep[e.g.][]{Smolen1991}, it should be mainly contained in the post-accretional global water ocean, thus contributing only weakly to the primitive atmosphere \citep[e.g.][]{HarmsWatzenberg1995}. Therefore, the conversion of NH$_3$ into N$_2$ should require efficient exchange processes from the global water ocean, the main reservoir of NH$_3$, to the atmosphere where dissociation processes occur. To determine for how long the conditions inducing the conversion of NH$_3$ into N$_2$ might have been met on early Titan, it is essential to better understand the exchange processes between the primitive atmosphere and the surface ocean. \\
The ocean-atmosphere exchanges on early Titan have also likely affected the carbon inventory, in particular CH$_4$ and CO$_2$ the two main carbon bearing species expected at the end of accretion \citep{Tobie2012}. Better understanding the partitioning of carbon species between the global ocean and the proto-atmosphere of Titan is essential to assess the most likely scenario for the origin of methane \citep{Atreya2010}. The isotopic $^{13}$C/$^{12}$C ratio in Titan's CH$_4$, measured by GCMS \citep{Niemann2010} is similar to CI carbonaceous chondrites, which suggests that it has been injected in Titan's atmosphere less than 1 Gyr ago \citep{Mandt2012}. As its lifetime in present-day atmosphere is of the order of $\sim$20-30 Ma due to irreversible photochemical processes and atmospheric escape in the upper atmosphere \citep{Wilson2004, Yelle2008, Krasnopolsky2010, Lavvas2011}, it must be replenished from a subsurface reservoir. The presence in the atmosphere of a significant amount of $^{40}$Ar, the decay product of $^{40}$K initially contained in the rocky core, indicates an efficient degassing from the interior \citep{Niemann2005, Niemann2010}, which may explain the CH$_4$ replenishment \citep{Tobie2006}. However, the nature and the location of the CH$_4$ subsurface reservoirs as well as their origin remain unconstrained. Methane could have been initially stored in the deep interior during the accretion \citep[e.g.][]{Tobie2006}, trapped during the post-accretional cooling stage \citep[e.g.][]{Lunine2009} or converted in the interior from CO$_2$ \citep[e.g.][]{Atreya2006,Owen2006, Glein2008}. In order to provide theoretical constraints on the origin of methane, we model the partitioning of the two main carbon species, CO$_2$ and CH$_4$, between the surface ocean and the primitive atmosphere during the early stage of Titan's history. \\
The amount of volatile compounds incorporated on Titan depends on the composition of planetesimals from which the Saturnian system formed and on the accretion processes within the Saturnian circumplanetary disk  \citep[e.g.][]{Mousis2009b}. \citet{Mandt2014} indicate that the composition of Titan's building blocks might be close to the cometary one. The cometary abundances range between 0.025 and 0.3 mol/mol H$_2$O for CO$_2$ (1.4 and 17 mol.kg$_{H_2O}^{-1}$), between 0.2$\times 10^{-2}$ and 0.015 mol/mol H$_2$O for NH$_3$ (0.1 and 0.83 mol.kg$_{H_2O}^{-1}$), between 0.0014 and 0.015 mol/mol H$_2$O for CH$_4$ (0.08 and 0.83 mol.kg$_{H_2O}^{-1}$) and between 0.004 and 0.3 mol/mol H$_2$O for CO  (0.22 and 17 mol.kg$_{H_2O}^{-1}$), that is also an abundant volatile found in comets \citep{Bockelee-Morvan2004, Mumma2011, Cochran2015}. Unfortunately, even if comets are representative of the initial composition of planetesimals in the vicinity of Saturn's forming region, some fraction of volatile compounds may have been lost during the accretion \citep[e.g.][]{Estrada2009} or converted in Titan's interior by water-silicate exchanges \citep{Matson2007}.\\ 
On the other hand, the latest estimate of volatile abundance in Enceladus plume derived from Cassini INMS provide an interesting alternative constraint for the volatile content of Titan: 0.3-0.8 \% (volume mixing ratio) of CO$_2$ (0.16-0.44 mol.kg$_{H_2O}^{-1}$); 0.4-1.3 \% NH$_3$ (0.22-0.72 mol.kg$_{H_2O}^{-1}$); 0.1-0.3 \% of CH$_4$ (0.06-0.17 mol.kg$_{H_2O}^{-1}$); 0.4-1.4 \% of H$_2$ (0.22-0.78 mol.kg$_{H_2O}^{-1}$) \citep{Waite2017}. We should, however, keep in mind that the volatile inventory in Enceladus have been re-processed due notably to aqueous alteration processes as suggested by the detection of H$_2$, and as a consequence may not be fully representative of the primordial composition \citep[e.g.][]{Zolotov2007}. In any case, it is worth noticing that CO$_2$ is a dominant species in both Enceladus' plume and comets.\\ 
Although CO$_2$ is predicted to be the main carbon species in models of Titan's formation \citep{Alibert2007, Hersant2008, Tobie2012}, its role in the chemical evolution of Titan has been ignored so far. This may be due to the fact that present CO$_2$ abundances are still unknown. Cassini detected only a small amount of CO$_2$ in Titan's upper atmosphere, most certainly originating from a recombination of CO and OH$^-$ \citep{Horst2008}. But it might be present in significant quantities on the surface, and potentially in the interior, as suggested both by the apparent CO$_2$ evaporation subsequent to the landing of the Huygens probe \citep{Niemann2010}, and the spectral absorptions indicative of CO$_2$ ice at different locations on Titan's surface \citep{McCord2008}. \\
CO$_2$ presence may have significantly affected the chemical evolution of the primitive atmosphere and ocean, the primitive cooling of the surface, as well as the formation of a primitive crust through clathration processes \citep[e.g.][]{Lunine2009, Choukroun2010}. Gas clathrate hydrates are crystalline water-based structures forming cages in which gas compounds such as CH$_4$ or CO$_2$ can be trapped \citep{Sloan2008, Choukroun2013}. It is then essential to determine in which conditions CH$_4$-CO$_2$ clathrates hydrates form and what were their composition and density during Titan's early history, in order to better understand the fate of the main carbon species during the post-accretional cooling stage and their potential impact on the subsequent evolution of Titan. \\
In this study, we investigate the partitioning of CO$_2$, CH$_4$ and NH$_3$ between the global water ocean of primitive Titan and its atmosphere, for variable initial volatile content. We also estimate the amount of volatile compounds that may be trapped in clathrates hydrates during the post-accretional cooling stage as well as the consequences for the formation of a primitive clathrate-rich crust. In section \ref{sec_model}, we detail the model developed to characterize the equilibrium between the atmosphere and a surface water ocean, including the description of the post-accretional structure of Titan, the gas-liquid equilibrium model and the clathration parameterization. In section \ref{sec_results}, the model results are displayed, with special focus on both the influence of the relative abundance of CO$_2$ and NH$_3$ on the ocean-atmosphere equilibrium, and the role of primordial CH$_4$ as well as N$_2$ resulting from NH$_3$ conversion on the clathration process.  Finally, we discuss in section \ref{sec_implications} the implications of our results for the conversion of NH$_3$ into N$_2$ and the trapping of CH$_4$ in the crust in Titan's primordial stages.\\

\section{Model}
\label{sec_model}

\subsection{Post-accretional structure of Titan}

\begin{figure}[h!]
\centering
\includegraphics[width=1\linewidth]{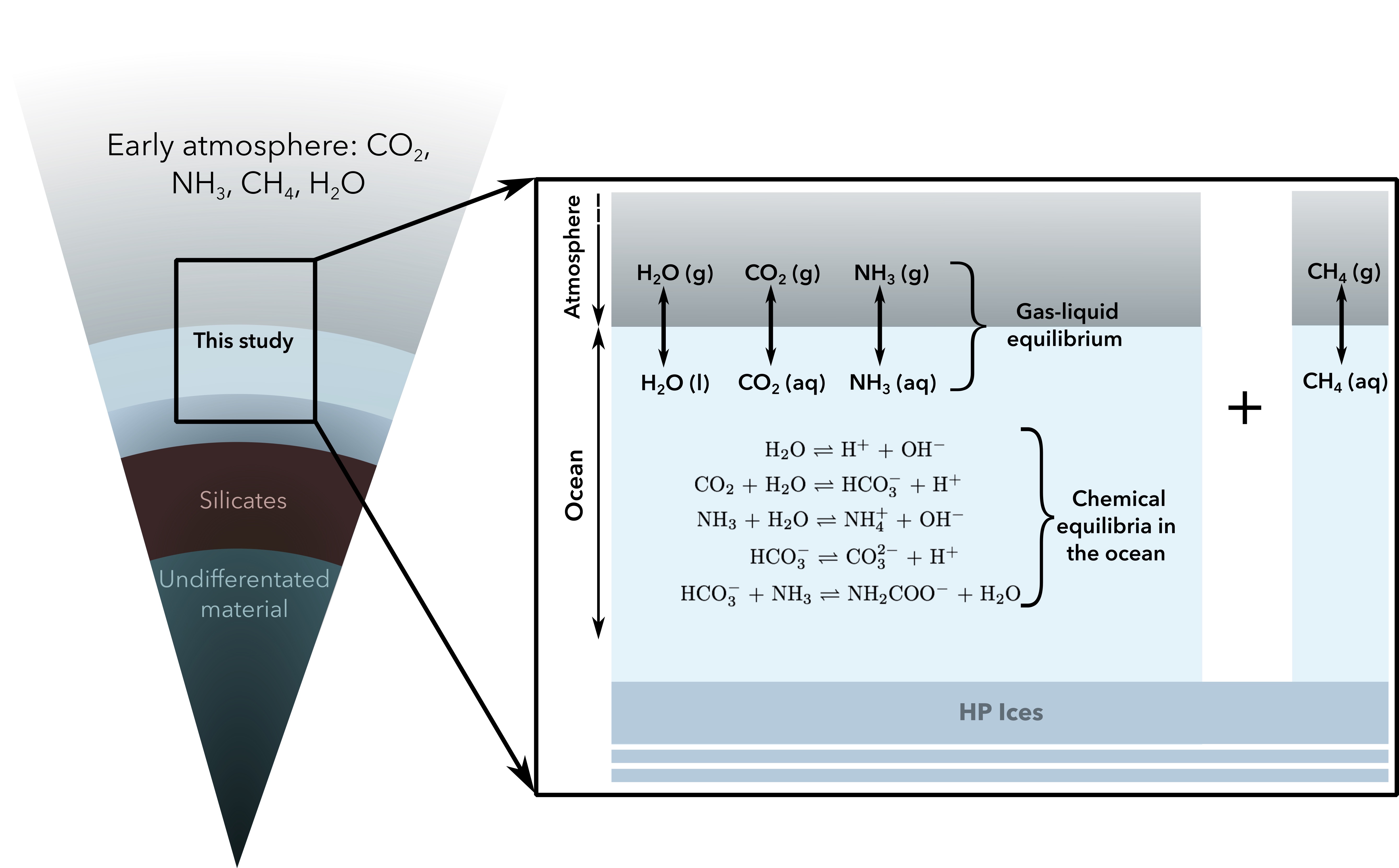}
\caption{\label{fig_struct}  On the left, the internal structure of primitive Titan (from \citet{Lunine1987}). On the right, the detailed sketch of the hydrosphere shows chemical interactions taken in account in our model. The separation between CH$_4$ and NH$_3$-CO$_2$ systems emphasizes the fact that no chemical interaction between CH$_4$ and NH$_3$-CO$_2$ in water is modeled, as explained in section \ref{subsec_vle}. Here $x_{melt}$ is set to 0.45, corresponding to undifferentiated core radius of 1195 km and a hydrosphere 853 km thick}
\end{figure}

To determine the amount of volatile species that are available to form Titan's primitive atmosphere at the end of the accretion, we consider a redistribution of the primordial material between the internal layers as shown in Figure \ref{fig_struct} and described in detail later. When Titan reaches a critical radius, which typical value varies between 750 and 1500 km, the impact heating during accretion becomes sufficient to melt the outer layer of the satellite \citep{Monteux2014}, thus allowing for the separation of water and volatiles and the deposit of silicates at the bottom of the global ocean. This could affect the overall chemical composition of volatiles due to chemical reactions between liquid water and silicates. Here we consider that this separation is fast enough that the chemical reactions do not modify considerably the initial volatile content of Titan's building blocks.

 In the model, the interior structure is divided in four main envelopes: an atmosphere, a water ocean, a high-pressure ice mantle and an ice-rock inner core and is sketched in Fig. \ref{fig_struct}. The mass ratio among the outer envelopes (atmosphere and ocean, made only of water and volatiles) and the inner parts (undifferentiated core + silicate mantle) is determined by the parameter x$_{melt}$ and the bulk mass fraction silicates of the satellite. Here we consider that Titan is formed from a mixture of 50 wt\% of silicates, and 50 wt\% of water and volatiles, \citet[e.g.][]{Lunine1987, Tobie2014}. x$_{melt}$ depends on the accretion timescale\citep[e.g.][]{Monteux2014} and was varied between 0.05 and 0.5 (the maximum value corresponding to all of the mixture water+volatiles melted during the accretion). Second, the bulk composition of the outer envelopes (i.e. the nature and the amount of volatiles compared to the amount of water) is fixed within a range consistent with what was discussed in section \ref{sec_intro}. Finally, one must chose a surface temperature at the interface between the surface ocean and the atmosphere, which can be understood as being an indicator of the time evolution. Our model is not time-dependent since each structure is at thermodynamic equilibrium. But a warmer surface temperature must be understood as being representative of an earlier stage in Titan’s evolution. Using these four inputs (surface temperature, primordial silicate fraction, primordial melt fraction, and bulk composition of the water-rich primordial material), it is then possible to describe both the size but also the composition of each sublayers of the outer part of the satellite (high-pressure icy layer, clathrate hydrate layer, liquid ocean, atmosphere).\\
 To determine the structure and composition at equilibrium for a given melt fraction, surface temperature, and bulk volatile content, the following iterative procedure is used. First, we assume that the outer parts of the satellite are made of an ocean and an atmosphere only. Then, for starting the iterative process, an initial surface pressure and an initial mass ratio between the vapor/atmosphere and the liquid/ocean phases are chosen arbitrarily. The existence of high-pressure ices and/or clathrates hydrates is then explored in depth (see sections 2.3 and 2.4 respectively). Depending on the ocean/ice interface depth, the total amount of volatiles that will be incorporated either in the ocean or the atmosphere is adjusted. By combining on one side the thermodynamic laws for L-V equilibrium (see section 2.2 for details), and the mass conservation constraints on the other side, one can find iteratively the unique solution of the problem for the initially chosen surface pressure, and the chosen mass ratio between ocean and atmosphere.  \\
 On the other hand, the surface pressure is constrained by the thickness and composition of the atmosphere. We assume an atmosphere that is isothermal, and uniform in composition, which implies that both pressure and density decreases exponentially with altitude with the same scale-height. In this condition, the surface pressure can be directly related to the total atmospheric mass (corresponding from the integration of the density profile from the surface to the space) (cf. eq. \ref{eq_Patm}). As most of the atmospheric mass is contained beneath the scale-height where the gravity varies only slightly with altitude, the exact surface pressure deviates only slightly from the value given by the following simple relationship:

\begin{equation}
P=\dfrac{M_{atm}g}{4 \pi R_T^2},
\label{eq_Patm}
\end{equation}

 where $g$ is the gravity at the surface of the satellite and $R_T$ is Titan's radius. By comparing the pressure obtained with equation (1) to the one chosen at the first step of this iterative process, the mass ratio between ocean and atmosphere can be adjusted. Then, the new pressure obtained from eq. \ref{eq_Patm} is chosen and the process is restarted from the determination of the HP ices and clathrate hydrate layers. At the end of this iterative process, one must find a unique solution as sketched in figure \ref{fig_struct}, where bulk composition and sizes of each layers are determined.\\

\subsection{Gas-liquid equilibrium model}
\label{subsec_vle}

At thermodynamic equilibrium in gas-liquid systems, fugacities $f_i$ of each component $i$ in both phases are equal: $f_i^{gas}=f_i^{liquid}$. This equality of fugacities are commonly expressed as: 

\begin{equation}
\phi_i y_i P = \gamma_i x_i f^0_i
\label{eq_base}
\end{equation}   
   
where $P$ is the total pressure, $y_i$ is the molar fraction of the component $i$ in the gas phase, $x_i$ is the molar fraction of the component $i$ in the liquid phase, $\phi_i(T, P, y_{i,j,k,...})$ is the fugacity coefficient, $\gamma_i(T, x_{i,j,k,...})$ the activity coefficient of $i$ in the aqueous phase, and $f^0_i$ is the reference state if $i$ in the liquid phase. Water is considered to be the only solvent and its reference state fugacity is its the fugacity of the pure component at system pressure and temperature (namely, the pressure of saturation at temperature T). The corresponding definition is not always possible for the solutes, unless one is willing to extrapolate the properties of some chemical species above their critical temperatures \citep[for a detailed description on the reference states see][]{Prausnitz1963, Prausnitz1998}. A more satisfactory procedure is to define the reference fugacity for solutes as follows:

\begin{equation}
f^o_i=\lim\limits_{x \rightarrow 0} \dfrac{f^{gas}_i}{x_i} = H_i,
\end{equation}

 where $H_i$ is the Henry's constant of $i$ in water. The Henry's constant has here the dimension of pressure. The expression of the water saturation vapor pressure as a function of temperature was taken from the NIST database \citep{NIST_webbook}, while the expression for the Henry's constant was taken from \citet{Rumpf1993co2} and \citet{Rumpf1993nh3} for CO$_2$ and NH$_3$, respectively.\\

 We use the Peng-Robinson-Gasem equation of state \citep{Peng1976, Gasem2001} to compute the fugacity coefficients of H$_2$O, CO$_2$ and NH$_3$ in the gas phase \citep{Englezos1993, Pazuki2006}. This is a semi-empirical approach that accounts for the non-zero volume of molecules and attractive forces between pairs of gas molecules. In the range of pressures and temperatures explored in this study, this approach is adequate for gas mixtures containing water but it requires sometimes to take into account the binary interaction coefficients for each pair of constituents. The comparison between the partial pressures in the gas phase predicted by our model and the experimental measurements for binary CO$_2$-H$_2$O and NH$_3$-H$_2$O systems \citep[e.g.][]{Smolen1991,Bamberger2000} indicates that binary interaction parameters are required only for the pair CO$_2$-H$_2$O, in which case the coefficient from \citep{Dhima1999} is used.\\
For the liquid phase and for the CO$_2$-NH$_3$-H$_2$O system, speciation equilibria occurring in the liquid water must necessarily be taken into account \citep{Edwards1975, Edwards1978, Thomsen2005}: 

\begin{equation}
CO_2+H_2O \overset{K_{CO_2}}{\rightleftharpoons} HCO_3^- + H^+,\\
\label{eq_chemeq1}
\end{equation}

\begin{equation}
NH_3+H_2O \overset{K_{NH_3}}{\rightleftharpoons}  NH_4^{+} + OH^{-},\\
\label{eq_chemeq2}
\end{equation}

\begin{equation}
HCO_3^{-} \overset{K_{HCO_3^-}}{\rightleftharpoons}  CO_3^{2-} + H^{+},\\
\label{eq_chemeq3}
\end{equation}

\begin{equation}
NH_3 + HCO_3^{-} \overset{K_{NH_2COO^-}}{\rightleftharpoons}  NH_2COO^{-}+H_2O,\\
\label{eq_chemeq4}
\end{equation}

\begin{equation}
H_2O \overset{K_{H_2O}}{\rightleftharpoons}  H^{+} + OH^{-}.
\label{eq_chemeq5}
\end{equation}

\begin{figure}[h!]
\centering
\includegraphics[width=0.85\linewidth]{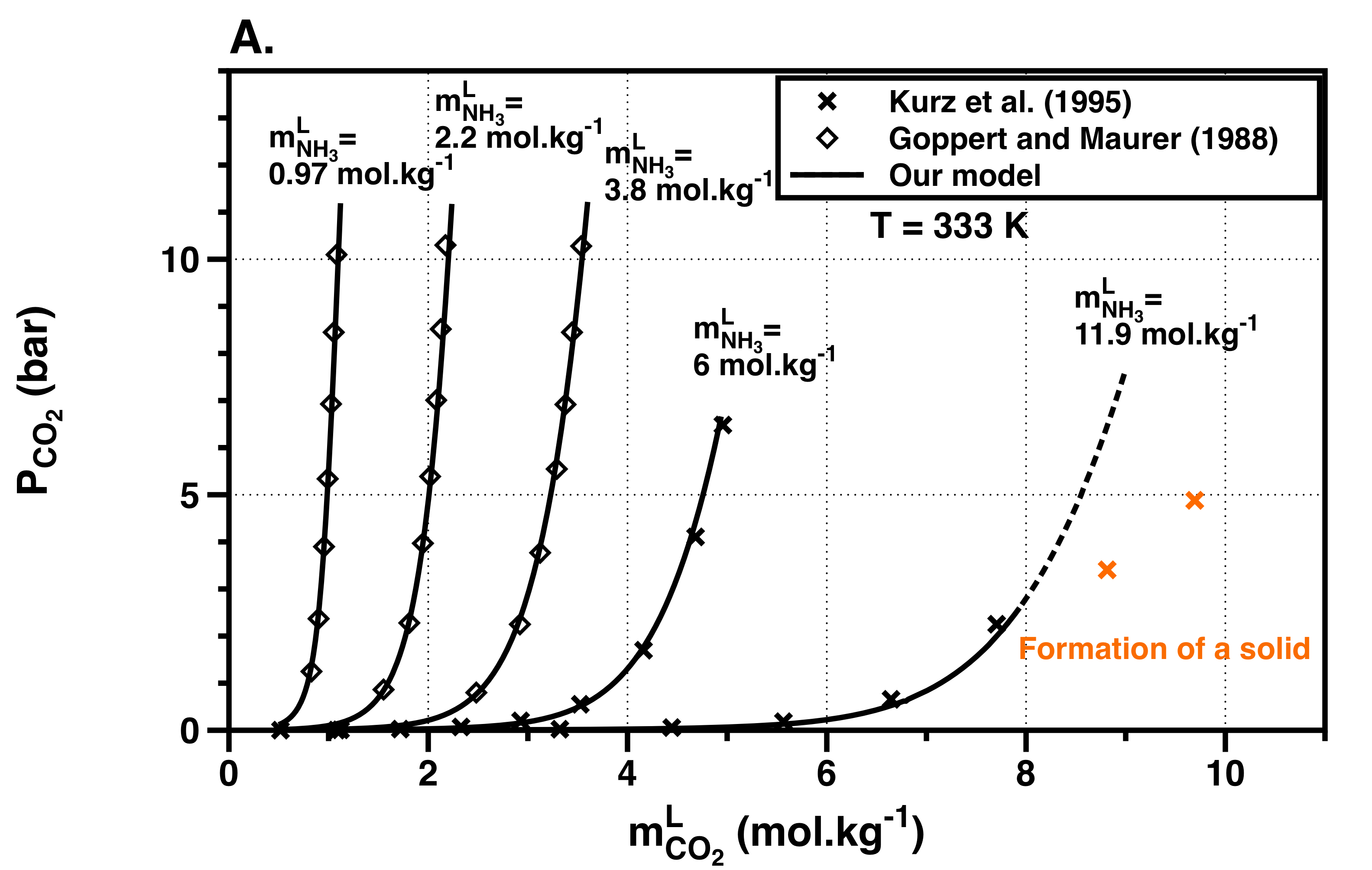}
\includegraphics[width=0.85\linewidth]{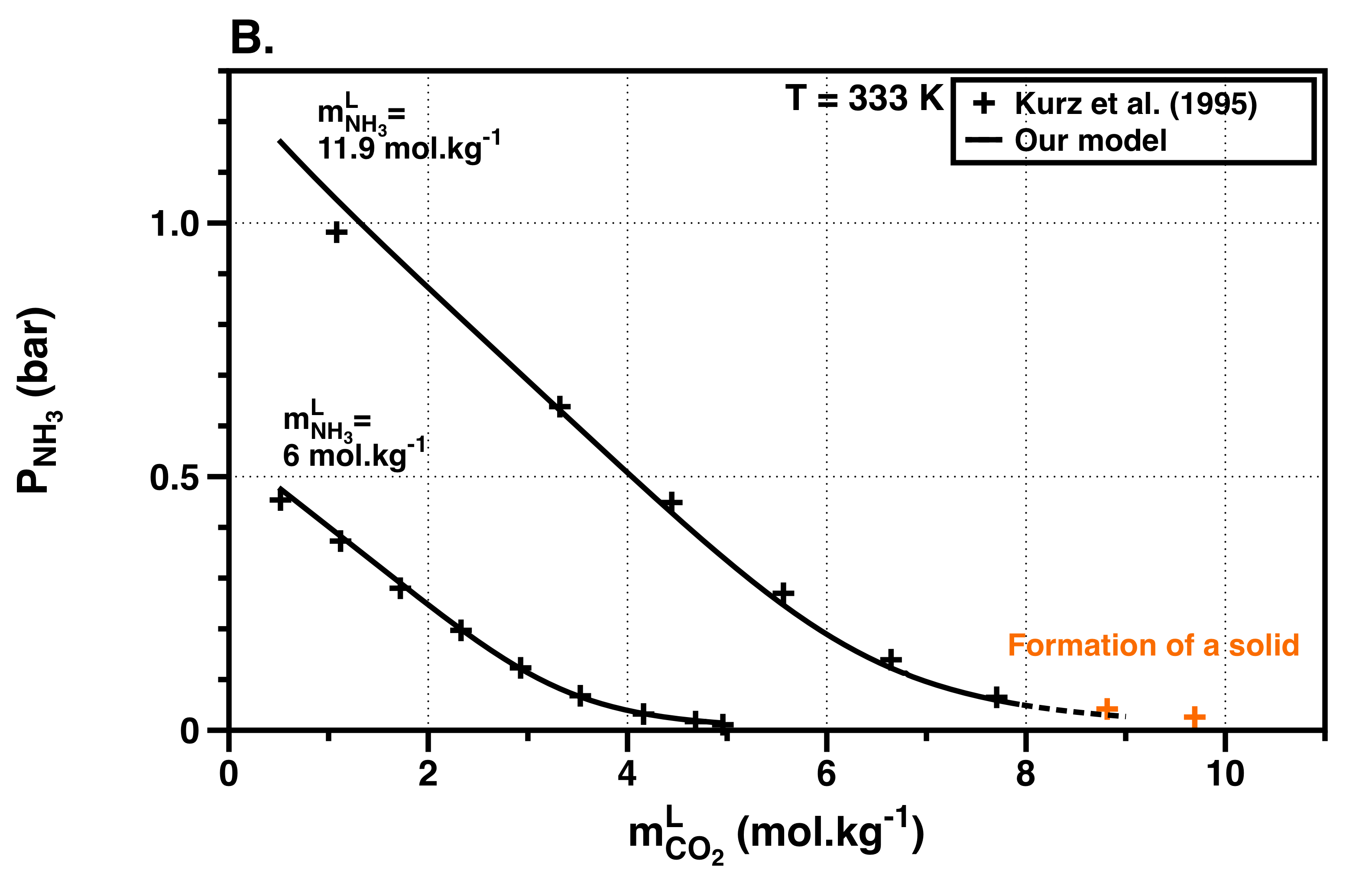}
\caption{ \label{fig_compexp} Comparison at T=333K of experimental (symbols) and computed (plain lines) partial pressures of CO$_2$ (A) and NH$_3$ (B) at equilibrium, as a function of CO$_2$ concentration in the liquid ($m^L_{CO_2}$), and for various concentration of NH$_3$ in the liquid ($m^L_{NH_3}$), for the CO$_2$-NH$_3$-H$_2$O system.}
\end{figure}

Dissolution of CO$_2$ in liquid water causes its partial dissociation and the formation of a weak acid, while the dissolution of NH$_3$ causes the formation of a weak base. Neglecting these dissociations for the ternary CO$_2$-NH$_3$-H$_2$O system induces large overestimations of the partial pressures in the gas phase. For the binary systems CO$_2$-H$_2$O and NH$_3$-H$_2$O, these reactions in water have almost no influence on the solubility of gases \citep[e.g.][for CO$_2$-H$_2$O system]{Spycher2003}. However for the ternary system CO$_2$-NH$_3$-H$_2$O, ionic species interact with each other and induce a significant increase in the amount of dissolved gases in water due to the formation of new ionic species, such as NH$_2$COO$^-$ \citep{Edwards1975, Edwards1978}. The equilibrium relations of the reactions described in Eq. \ref{eq_chemeq1} to \ref{eq_chemeq5} can be written: 

  \begin{equation}
   \begin{split}
     K_{CO_2}=&\dfrac{ \gamma_{HCO_3^-} \gamma_{H^+} }{ \gamma_{CO_2} \gamma_{H_2O} } \dfrac{ m_{HCO_3^-} m_{H^+}}{ m_{CO_2} m_{H_2O}}, \\   
    \end{split}
     \label{eq_Kdiss1}
    \end{equation}

      \begin{equation}
      \begin{split}
      K_{NH_3}=&\dfrac{ \gamma_{NH_4^+} \gamma_{OH^-} } {\gamma_{NH_3} \gamma_{H_2O}}  \dfrac{ m_{NH_4^+} m_{OH^-} }{  m_{NH_3} x_{H_2O}  }, \\
        \end{split}
       \label{eq_Kdiss2}
        \end{equation}

       \begin{equation}
       \begin{split}
      K_{HCO_3^-}=&\dfrac{ \gamma_{H^+}  \gamma_{CO_3^{2-}} }{ \gamma_{HCO_3^-} } \dfrac{ m_{H^+}  m_{CO_3^{2-}} }{  m_{HCO_3^-} }, \\
        \end{split}
       \label{eq_Kdiss3}
       \end{equation}
    
       \begin{equation}
       \begin{split}
      K_{NH_2COO^-}=&\dfrac{ \gamma_{H_2O} \gamma_{NH_2COO^-} }{ \gamma_{NH_3} \gamma_{HCO_3^-}  } \dfrac{ m_{H_2O} m_{NH_2COO^-} }{ m_{NH_3} m_{HCO_3^-} },\\
        \end{split}
        \label{eq_Kdiss4}
        \end{equation}
    
        \begin{equation}
        \begin{split}
      K_{H_2O}=&\dfrac{ \gamma_{H^+} \gamma_{OH^-} }{ \gamma_{H_2O}}  \dfrac{m_{H^+} m_{OH^-} }{ x_{H_2O} },
       \end{split}
       \label{eq_Kdiss5}
       \end{equation}
    
where $m_i$ represent the molality of the indicated compound in water (i.e. the number of mol of the chemical compound normalized to 1 kg of water, units: mol.kg$^{-1}$) and $K_i$ are the dissociation coefficients, which depend on temperature only. The model is extremely sensitive to the values of K$_i$. The best correspondence between experimental data (see the compiled list the experimental points of CO$_2$-NH$_3$-H$_2$O system in \citet{Darde2010}) and our model was obtained by using the parameters from \citet{Darde2010}.  In this study, K$_i$ are considered to be function of temperature only. The parameters proposed by \citet{Darde2010} are valid up to 100 bar and we do not cross this limit. \\
The presence of ionic species makes the computation of activity coefficients $\gamma_i$ for the chemical species in the liquid phase (both for water and dissolved species) mandatory. To estimate these activity coefficients, we use the extended UNIQUAC model (short for extended Universal QUAsi-Chemical model) \citep{Anderson1978, Sander1986, Nicolaisen1993, Thomsen1999} with parameters taken from \citet{Darde2010}. For the ternary system CO$_2$-NH$_3$-H$_2$O, the state of equilibrium is obtained by solving simultaneously the equilibrium relations for the dissolved species (Eq. \ref{eq_Kdiss1} to \ref{eq_Kdiss5}), mass balance, charge balance and gas-liquid equilibrium for the non-ionic species: CO$_2$, NH$_3$ and H$_2$O (Eq. \ref{eq_base}). \\ 
Our comparisons between experimental data listed in \citet{Darde2010} and the model are similar to those obtained by \citet{Darde2010}, who reproduced the CO$_2$ and NH$_3$ partial pressure ($P_{CO_2}$, $P_{NH_3}$) with mean deviation of 11\% and 12\%, respectively (see Fig. \ref{fig_compexp}). \\
The partial pressure of water is almost not affected by the dissociation reactions that occur in the liquid and remains close to its saturation vapor pressure. Deviation to this value occurs only for high concentration of CO$_2$ \textit{and} NH$_3$, typically $m_{CO_2}$ and $m_{NH_3}$ larger than 10 mol.kg$^{-1}$ \citep{Goppert1988}, that corresponds to values  up to $\sim$10 times larger than the one expected on primitive Titan for NH$_3$.  Moreover, for high CO$_2$ and NH$_3$ concentrations, the formation of solid phases (NH$_4$HCO$_3$, NH$_4$COONH$_4$, (NH$_4$)$_2$CO$_3$H$_2$O or (NH$_4$)$_2$CO$_3$2NH$_4$ HCO$_3$) was observed experimentally \citep[c.f.][]{Kurz1995, Jilvero2015}. The composition of the precipitated solid depends on the relative abundances of CO$_2$ and NH$_3$ \citep{Thomsen1999, Darde2010, Darde2012} and it formation lead to a decrease of the CO$_2$ partial pressure in the gas phase. As we do not account for the formation of the solid phases, our model is valid for (i) concentrations up to 6 mol.kg$^{-1}$ for NH$_3$ and up to 4 mol.kg$^{-1}$ for CO$_2$ (see Fig. \ref{fig_compexp}) and (ii) temperatures between 273.15 and 373.15 K.  Considering the limitations in volatile concentrations of our gas-liquid equilibrium model and the volatile abundances observed in comets and in Enceladus' plume, we will explore the range of concentrations up to 4 mol.kg$^{-1}$ for CO$_2$ and up to 1 mol.kg$^{-1}$ for NH$_3$, which is in between the validity range of our domain. \\

  To model the dissolution of CH$_4$ along with CO$_2$ and NH$_3$ in water using the extended UNIQUAC model, one has to fit the experimental data to retrieve the binary interaction coefficients. However, only limited experimental data exists for the CO$_2$-CH$_4$-H$_2$O system below 373K, the upper temperature limit that is relevant for this study \citep{AlGhafri2014}. To our knowledge, no experimental data are available for the gas-liquid equilibrium of CH$_4$-NH$_3$-H$_2$O and CH$_4$-CO$_2$-NH$_3$-H$_2$O systems. Consequently, we model the dissolution of CH$_4$ using the Henry's law and therefore we neglect the interactions between CH$_4$ and (CO$_2$, NH$_3$) and the resulting ionic species in water. The CH$_4$-H$_2$O Henry's constant was obtained by a fit of the experimental data of \citet{Cramer1984}: 
  
  \begin{equation}
  H_{(CH_4,H_2O)}=2.9477\times10^6-44139T+246.83T^2-0.64697T^3+8.0669\times10^{-4}T^4-3.8742\times10^{-7},
  \end{equation}
    
    where the Henry's law constant is in bar and the temperature in kelvin.\\
    The comparison between the computed CH$_4$ partial pressures in the CO$_2$-CH$_4$-H$_2$O system using the Henry's constant and the experimental data for the same system at T=323.15K \citep{AlGhafri2014} shows a difference up to 12 bar. This deviation is greater for lower pressures, leading to an error up to 100\%. This error decreases with the increasing pressure.\\

    \subsection{Formation of high-pressure ices}    
    
   Depending on the surface oceanic temperature and on the thickness and the thermal structure of the ocean, a high-pressure (HP) ice layer may form in between the ocean and the silicate mantle (\ref{fig_struct}).  Assuming an adiabatic pressure-temperature profile in pure water system, the HP ice layer forms only if the surface temperature is colder than 291 K. Its higher thickness is about 180 km when the surface temperature is down to the freezing point of pure water. Neglecting the possible formation of clathrates hydrates, the position of the HP icy layer-ocean interface is fixed from the intersection between the adiabatic temperature profile in the ocean and the melting curve of ice VI \citep{Choukroun2007}. The melting temperature depends on pressure, but also on the amount of volatiles dissolved in the liquid layer.  For sake of simplicity, we neglect the effect of the dissolved salts and volatile compounds on the melting curve, a reasonable assumption since the volatile content does not exceed a few percents, and we use the pure water ice VI melting curve in this model. This implies a small underestimate of the thickness of the water reservoir, as well as an overestimate of the volatile contents in the liquid. This effect is considered negligible when compared to the degree of uncertainty on the bulk composition of Titan's building blocks. \\

 \subsection{Formation of clathrate hydrates}   
Carbon dioxide and methane can also be mixed with water in clathrate hydrate solid phases (refererred simply as clathrates for the rest of the manuscript), with a relative proportion that is controlled by the surface temperature and their partial pressure in the atmosphere. As we also explore the influence of nitrogen on the formation of clathrates, we need to account for the incorporation of N$_2$ in the CO$_2$-CH$_4$ clathrate. In order to determine the formation temperature and the composition of the clathrate hydrates for a mixture of gases, we use the simplified modeling approach of \citet{Lipenkov2001}. We used a RMS analysis to fit the experimental data for liquid water-hydrate-gas equilibrium of CH$_4$, CO$_2$ or N$_2$ between 273.15K and 283.15K \citep{Sloan2008} to a third degree polynomial:

\begin{equation}
P_{d,i} = AT^3+BT^2+CT+D, 
\end{equation} 

where $P_{d,i}$ is the pressure of formation of a pure clathrate (CH$_4$, CO$_2$ or N$_2$) and coefficients A, B, C, and D are summed up in Table \ref{tab_clathrates}. 

\begin{table}[h!]
\centering
\begin{tabular}{c|cccc}
\hline 
Component & $A$ (bar.K$^{-3}$) & $B$ (bar.K$^{-2}$) & $C$ (bar.K$^{-1}$) & $D$ (bar) \\ 
\hline 
\hline 
CH$_4$  & -3.1327$\times$10$^{-4}$     & 0.511707   & -207.353     & 2.4870$\times$10$^4$  \\ 
CO$_2$  & 1.062142$\times$10$^{-2}$  & -8.63639    & 2342.1086  & -2.11830$\times$10$^5$  \\ 
N$_2$    & 8.48933$\times$10$^{-2}$    & -70.17086  & 19359.66    & -1.78255$\times$10$^6$  \\  
\end{tabular} 
\label{tab_clathrates}
\end{table}

These stability curves are derived from experiments for single gas compound. They can be used to approximate the stability curve of a multiple compound gas hydrate at first order, by assuming that clathrate hydrate behaves as an ideal solution and that the ratio of occupancies for the small and large cages is constant and equal for all guest molecules \citep[e.g][]{Miller1974, Lipenkov2001, McKay2003}. Within these assumptions, the formation pressure of multiple compound clathrates $P_d$ is predicted using the following relationship: 

\begin{equation}
\left(P_d \right)^{-1} = \sum_i \left[ \dfrac{y_{c,i}}{P_{d,i}} \right],
\label{eq_mixclath}
\end{equation}

where $y_{c,i}$ is the mole fraction in the gas mixture that account only for species that are trapped in clathrates (NH$_3$ and H$_2$O gases are not considered for $y_{c,i}$ computation).\\ 
For a given surface temperature, the clathrate formation pressure is computed using the atmospheric composition that is determined with the gas-liquid equilibrium model described in section \ref{subsec_vle}. If the predicted clathrate formation pressure is equal to or exceeds the total surface pressure, clathrate hydrates form with a composition that is fixed from the following relationship: 

\begin{equation}
\dfrac{X_i P_{d,i}}{X_j P_{d,j}} = \dfrac{f_i}{f_j},
\label{eq_mixclath_comp}
\end{equation}

where $X_i$ and $X_j$ are the mole fraction of gases $i$ and $j$ in the clathrate phase, $f_i$ and $f_j$ are fugacities in liquid water (right-hand term of Eq. \ref{eq_base}). This approach has been used to determine the clathrate formation in the Lake Vostok \citep{Lipenkov2001, McKay2003} or the formation of CH$_4$ clathrates in the marine sediments. The computations of $P_d$ using the equation \ref{eq_mixclath} give nearly the same results as those obtained from more rigorous thermodynamic models \citep{Lipenkov2001, Choukroun2013a}. \\ 
In order to determine whether the formed clathrate hydrate would remain stable at the ocean surface or sink to the oceanic floor, we compute its density $\rho$ using the following formula \citep{Sloan2008}: 

\begin{equation}
\rho = \dfrac{ N_w \times M_{H_2O} + \sum_{i=1}^C \sum_{j=1}^N y_{ij} \nu_i M_j }{ N_A \times V_{cell} }
\label{eq_rho}
\end{equation}

Where $N_w$ is the number of water molecules per unit cell (46 for structure sI that is considered here), $N_A$ is the Avogadro number, $V_{cell}$ is the volume of unit cell (12 $\AA^3$ for the sI structure), M$_{H_2O}$ is the molecular weight of H$_2$O, M$_j$ is the molecular weight of component $j$, $y_{ij}$ is the fractional occupation of cavity $i$ by component $j$, assumed to be equivalent in small and large cages and equal to $X_j$ (corresponding to full occupancy), $\nu_i$ is the number of type $i$ cavities per water molecule in unit cell (2 and 6 for small and large cages respectively), $N$ is the number of cavities in the unit cell and $C$ is the total number of compounds trapped in hydrate phase.\\

\section{Results}
\label{sec_results}
In this section, we explore various primordial composition characterized by different values of $m_{CO_2}$/$m_{NH_3}$ molar ratio. We first outline the general behavior of the CO$_2$-NH$_3$-H$_2$O system and define two compositions that are relevant for the atmosphere of early Titan. In the second part, the effect of the surface temperature on the gas-liquid equilibrium is explored for these two cases. Finally, methane and diatomic nitrogen are included to investigate the conditions for formation of clathrate hydrates during Titan's primordial history. 

\subsection{The effect of the $m_{CO_2}$/$m_{NH_3}$ molar ratio in the CO$_2$-NH$_3$-H$_2$O system}
\label{subsec_molalityeffect}
\begin{figure}[h!]
\centering
\includegraphics[width=0.75\linewidth]{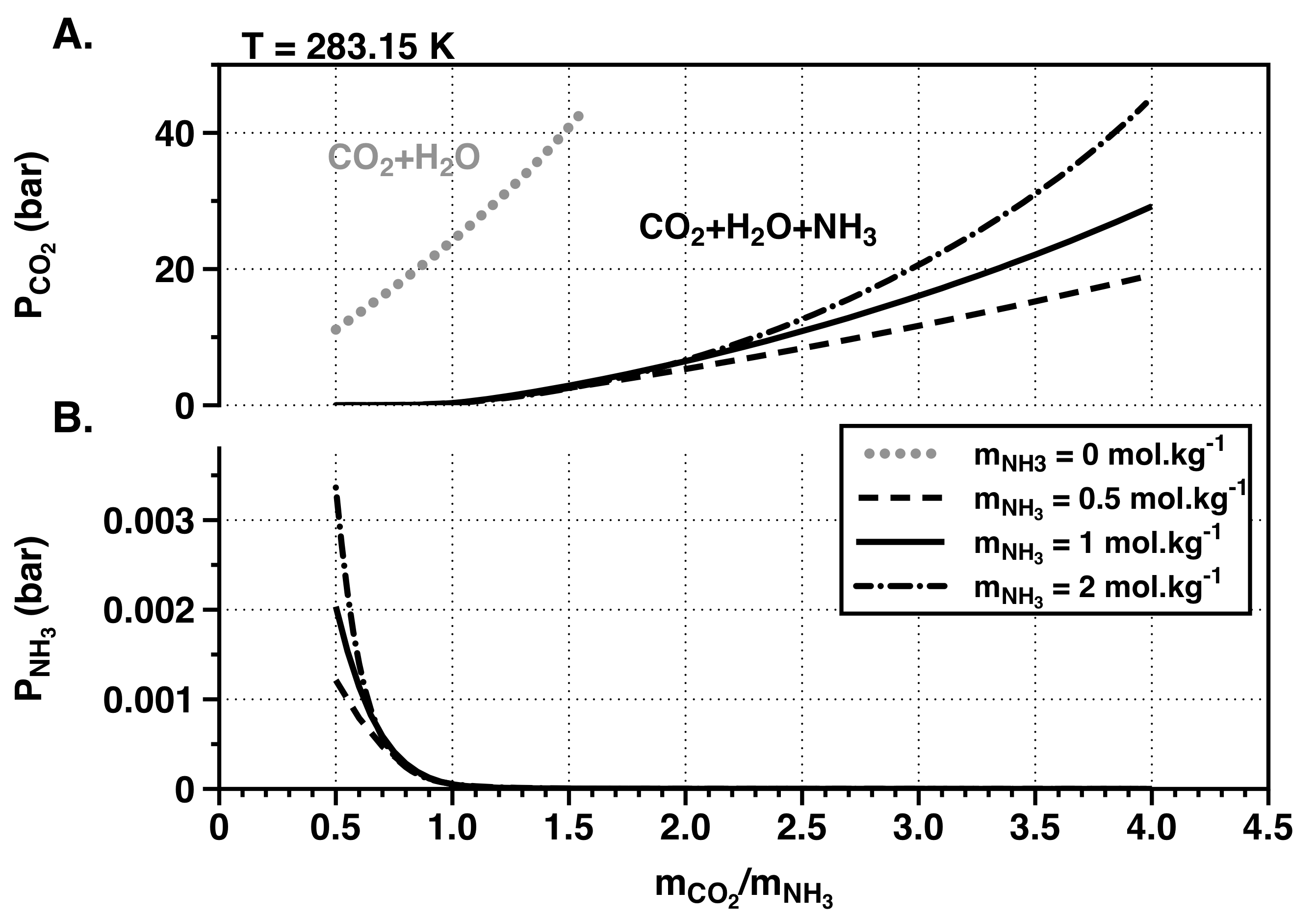}
\includegraphics[width=0.75\linewidth]{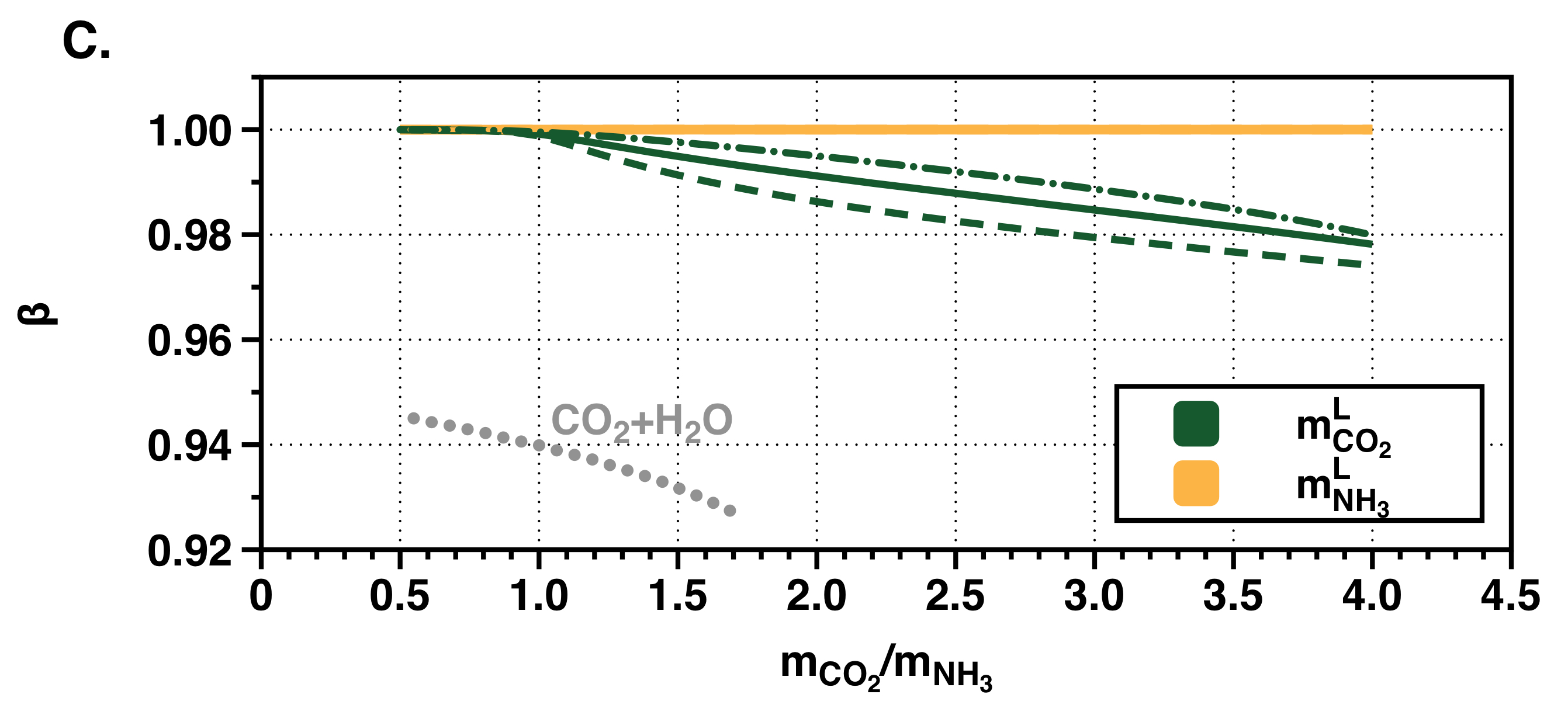}
\caption{ \label{fig_vleresults} Partial pressures in the gas phase of CO$_2$ (A) and NH$_3$ (B) $m_{CO_2}$/$m_{NH_3}$ molar ratio. In Figure (C), the molar fraction $\beta$ of CO$_2$ (or NH$_3$) that is dissolved in the liquid phase respectively to the global numbers of moles of CO$_2$ (or NH$_3$) in the hydrosphere+atmosphere is shown. Figure (C) illustrates well the fact that ammonia is almost entirely dissolved in the liquid phase. In panels (A) and (C), we show for comparison the partial pressure of CO$_2$ and the $\beta$ fraction for the CO$_2$-H$_2$O binary system. The values are displayed as a function of the total CO$_2$ molality.}
\end{figure}

Figure \ref{fig_vleresults} summarizes the general behavior of the gas-liquid equilibrium in the CO$_2$-NH$_3$-H$_2$O system for a temperature of 283.15 K. It displays the repartition of CO$_2$ and NH$_3$ in both liquid and gas phases as a function of the ratio of molalities $m_{CO_2}$/$m_{NH_3}$. Two distinct cases can be identified depending on the fact that NH$_3$ is the dominant compound of the system ($m_{CO_2}$/$m_{NH_3}<$1), or CO$_2$ is the main compound ($m_{CO_2}$/$m_{NH_3} >$1) (Fig. \ref{fig_vleresults} A and B). When CO$_2$ is less abundant than NH$_3$, the partial pressure of CO$_2$ in the gas phase remains low, because the equilibrium reactions (eq. \ref{eq_chemeq1} to \ref{eq_chemeq5}) in liquid water trap almost all the carbon dioxide in the liquid phase. When CO$_2$ is more abundant than NH$_3$, the partial pressure of CO$_2$ in the gas phase increases with its molality in the ocean + atmosphere system. Figure \ref{fig_vleresults} C displays $\beta$, the number of mols of CO$_2$ dissolved in water relative to the total number of moles of CO$_2$ in the ocean + atmosphere system, as a function of $m_{CO_2}$/$m_{NH_3}$ concentration ratio. It shows that, for the ternary CO$_2$-NH$_3$-H$_2$O system, more than 97\% of CO$_2$ available in the hydrosphere is dissolved in water. For a case without NH$_3$, this fraction is systematically lower than 95\%, as displayed in Figure \ref{fig_vleresults} C, and the CO$_2$ partial pressure in the atmosphere is higher (see Fig. \ref{fig_vleresults} A).\\
On the other hand, ammonia is almost entirely dissolved in water. The ammonia molality in the ocean $m^{L}_{NH_3}$, is almost equal to its global molality $m_{NH_3}$. The NH$_3$ partial pressure in the gas phase is several orders of magnitude lower than the partial pressure of CO$_2$ and is always decreasing with the CO$_2$ increasing global molality. This trend will be observed at all temperatures explored in this study, and is in agreement with previous experimental works \citep[see][]{VanKrevelen1949, Verbrugge1979, Muller1988, Goppert1988, Kurz1995, Jilvero2015}.  

\subsection{Effect of the temperature and the presence of high pressure ices in the CO$_2$-NH$_3$-H$_2$O mixture for primitive Titan}
\begin{figure}[h!]
\centering
\includegraphics[width=0.75\linewidth]{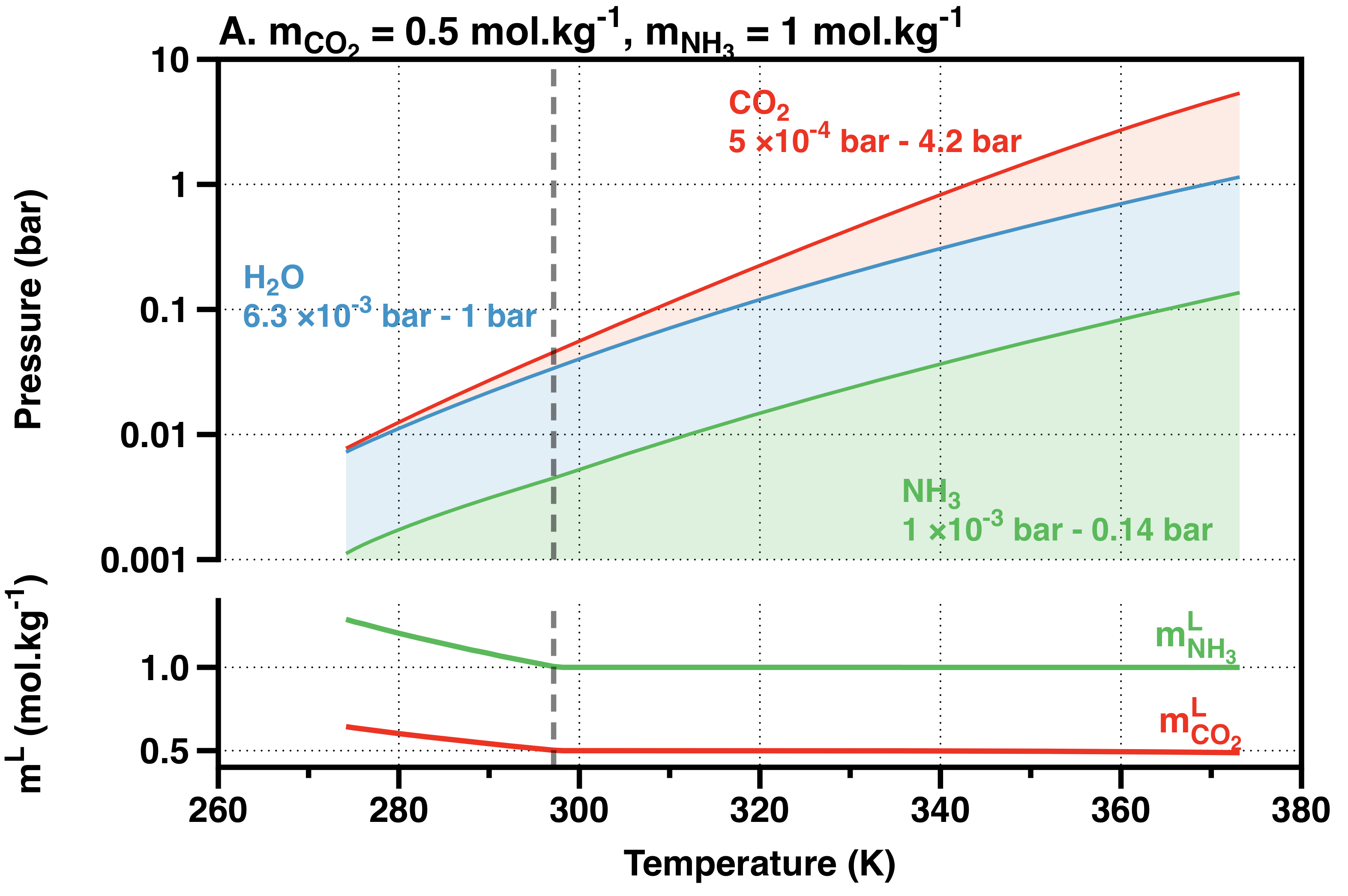}
\includegraphics[width=0.75\linewidth]{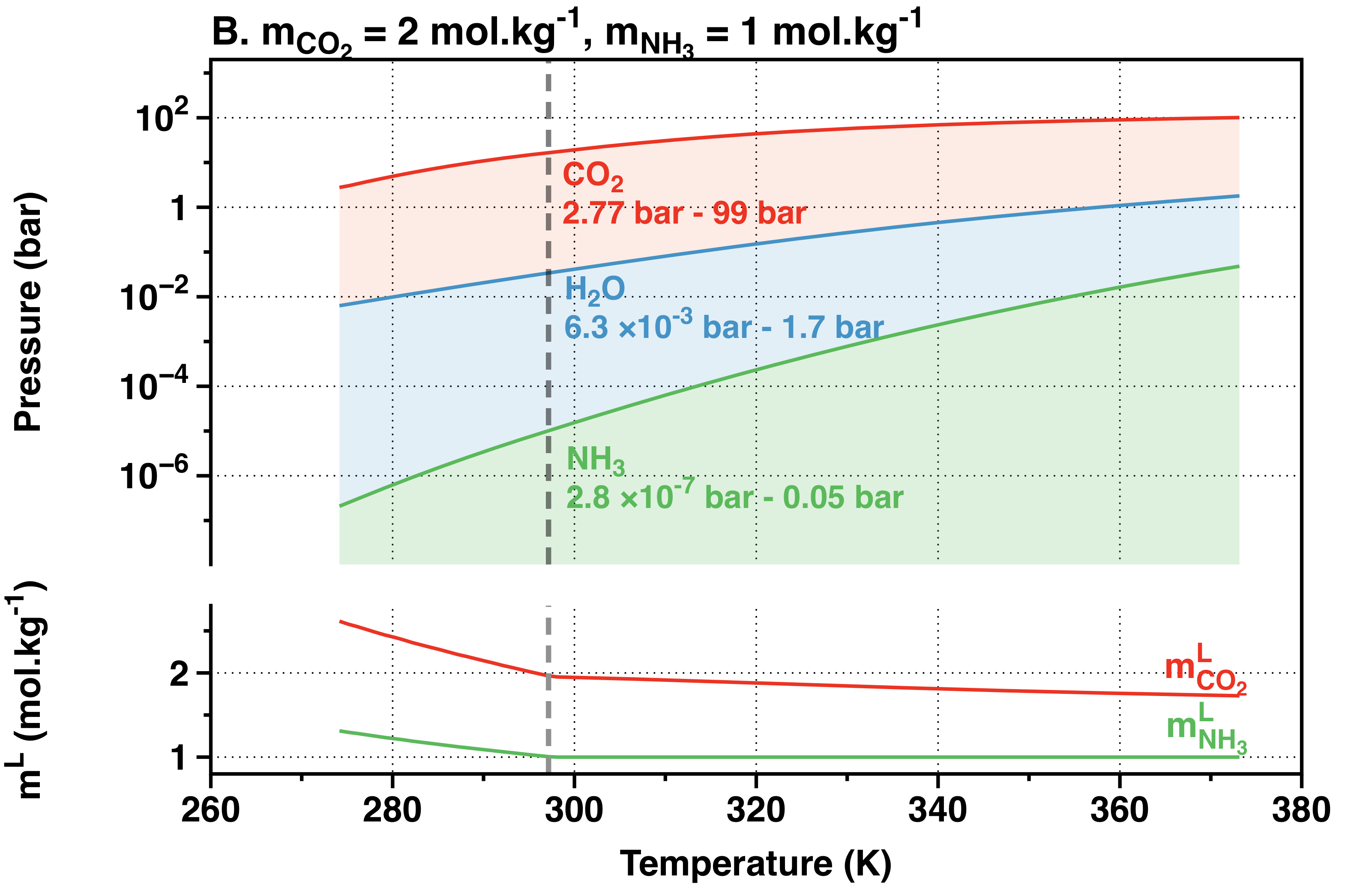}
\caption{\label{fig_resPL3} Evolution of the total pressure and atmospheric composition (upper part) and of the molalities in the liquid phase (bottom part) as a function of temperature for $m_{NH_3}$ = 1 mol.kg$^{-1}$ and $m_{CO_2}$ = 0.5 mol.kg$^{-1}$ (A), or $m_{CO_2}$ = 2 mol.kg$^{-1}$ (B). The numbers in colors indicate the minimal (T=273 K) and maximal (T=373 K) partial pressures for each chemical species. The vertical grey line indicates the onset temperature below which ice VI is formed at the oceanic floor in these simulations. The colored areas (blue, green and red) show the contribution of each gas (water, ammonia and carbon dioxide, respectively) to the total pressure: e.g. the level of P$_{NH_3}$ is indicated with the green curve and the level of P$_{H_2O}$ is the difference between the blue and the green curves. The red curve corresponds to the total pressure (P$_{CO_2}$+P$_{NH_3}$+P$_{H_2O}$).} 
\end{figure}

For this set of calculations with varying surface temperature of the global water ocean, the NH$_3$ global concentration is set to 1 mol.kg$^{-1}$, and the concentration of CO$_2$ is fixed either to 0.5 or 2 mol.kg$^{-1}$, in order to consider the two regimes described above. In both cases, the composition of the gas phase will be explored throughout the entire possible temperature range (273-373 K) of the gas-liquid interface. This temperature variation could be assimilated to a time evolution from early stages (high temperature) to later stages prior to the upper crust formation (low temperature) on early Titan's surface. The position of the interface between ice VI and ocean must also vary with the surface temperature (see details below), as the intersection between the ocean adiabatic curve and the melting curve shifts upward with decreasing surface temperature. \\
Figure \ref{fig_resPL3} shows the evolution of the total atmospheric pressure at all surface temperatures, and the contribution of each gas to this total pressure. For all cases, NH$_3$ is mainly dissolved in water, and its partial pressure in the atmosphere remains low with respect to the two other compounds. The atmospheric pressure of NH$_3$ decreases with increasing concentration in CO$_2$ (Fig. \ref{fig_resPL3} A and B), as already demonstrated in Figure \ref{fig_vleresults}. As for CO$_2$, it was shown previously that its partial pressure varies significantly depending on the relative abundance of CO$_2$ and NH$_3$. For $m_{CO_2}$/$m_{NH_3}<$ 1, the partial pressure of CO$_2$ in the atmosphere remains low and two regimes can be identified: at low temperatures ($<$323 K in this case), water vapor is the main atmospheric component; at high temperatures ($>$323 K in this case), CO$_2$ is the major component in the atmosphere (see Fig. \ref{fig_resPL3} A). When $m_{CO_2}$/$m_{NH_3}>$ 1 (Fig. \ref{fig_resPL3} B), the main atmospheric component is the carbon dioxide, whatever the temperature is, and water vapor is about two order of magnitude less abundant. \\
Finally, we find that the partial pressure of water is weakly affected by the dissociation reactions that occur in the liquid and remains close to the saturation vapor pressure at all temperatures and all compositions. \\

In this simulation we used $x_{melt}$=0.5, possibly the highest value that could be envisaged when considering that Titan is mostly half water, half silicates by mass. Within this assumption, it turns out that ice VI starts to form on the oceanic floor for temperatures lower than T=297K. Since it is assumed that ice VI is made of pure water (i.e. neglecting the small amount of volatiles that could be incorporated in the ice and/or possible hydrates), the amount of volatiles in the liquid layer is increased significantly as the HP ice layer thickens (see bottom panels of Fig. \ref{fig_resPL3} A and B). Furthermore, the increase of ammonia concentration in liquid water imposes a higher solubility of carbon dioxide because of the dissociation reactions in water, as explained above. It leads to an accentuated decrease of the partial pressures of these gases in the atmosphere at temperatures where ice VI could be formed. \\

\subsection{Formation of methane-rich clathrates in the CH$_4$ - CO$_2$ - NH$_3$ - N$_2$ - H$_2$O system}
To investigate the formation of clathrate hydrates, we accounted for the presence of CH$_4$ and N$_2$ in CO$_2$-NH$_3$-H$_2$O system. The effect of the methane is investigated by using the Henry's law (see section \ref{subsec_vle}), which limits its validity to cases where the interaction of dissolved methane with other compounds in solution can be neglected. In principle, this does not mean that the amount of methane must be very small since it is not very reactive (methane is neither polar nor reactive, not very soluble, and it does not dissociate in water. Nonetheless, since it influences the solubility of other compounds \citep[see][]{AlGhafri2014}, we suggest to restrain our investigations to small amounts of methane where possible interactions should be limited. A fixed CH$_4$ concentration of 0.2 mol.kg$^{-1}$ has been chosen for illustrative purposes. Results are very preliminary, but illustrate well what are the two different regimes that could have existed on early Titan. \\
First of all, it must be reminded that the solubility of CH$_4$ in water is extremely low, thus implying that the atmosphere constitutes the main reservoir of CH$_4$. Consequently, a first regime can be easily described. Indeed, if $m_{CO_2} < m_{NH_3}$, it has been shown in section \ref{subsec_molalityeffect} that CO$_2$ is efficiently dissolved in water and its partial pressure in the atmosphere remains low. Thus, the main gas in the atmosphere is methane. Therefore, when the total surface pressure reaches the pressure of formation of clathrates, these clathrates are mostly made of CH$_4$ and are consequently less dense that the oceanic water. It turns out that a crust must be formed rapidly at the interface between the ocean and the methane-rich primordial atmosphere, thus limiting exchanges between the two reservoirs. The question is then to determine the critical amount of carbon dioxide above which another evolution involving clahtrate sinking could be envisaged.\\

\begin{figure}[h!]
\centering
\includegraphics[width=0.65\linewidth]{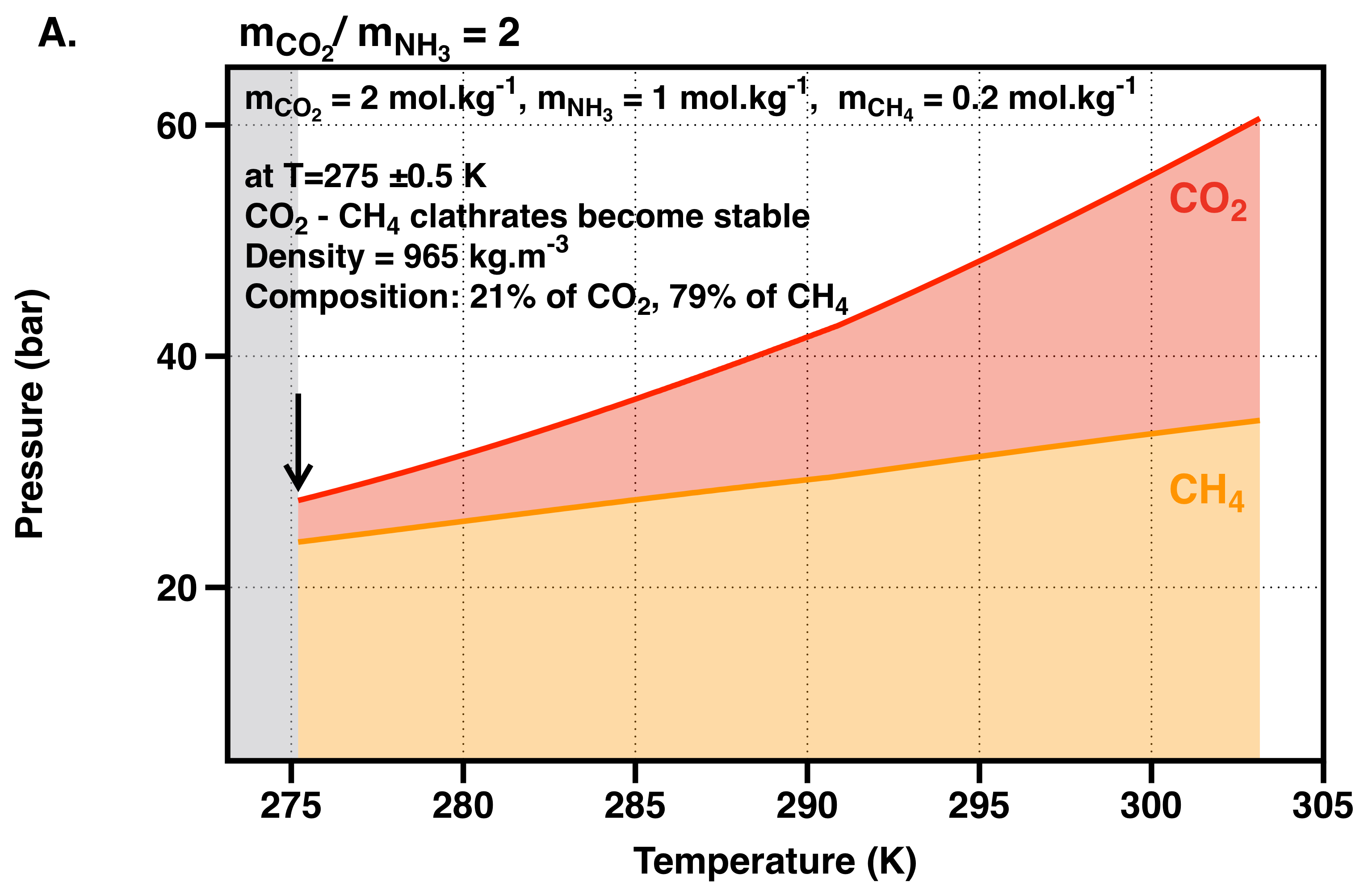}
\includegraphics[width=0.65\linewidth]{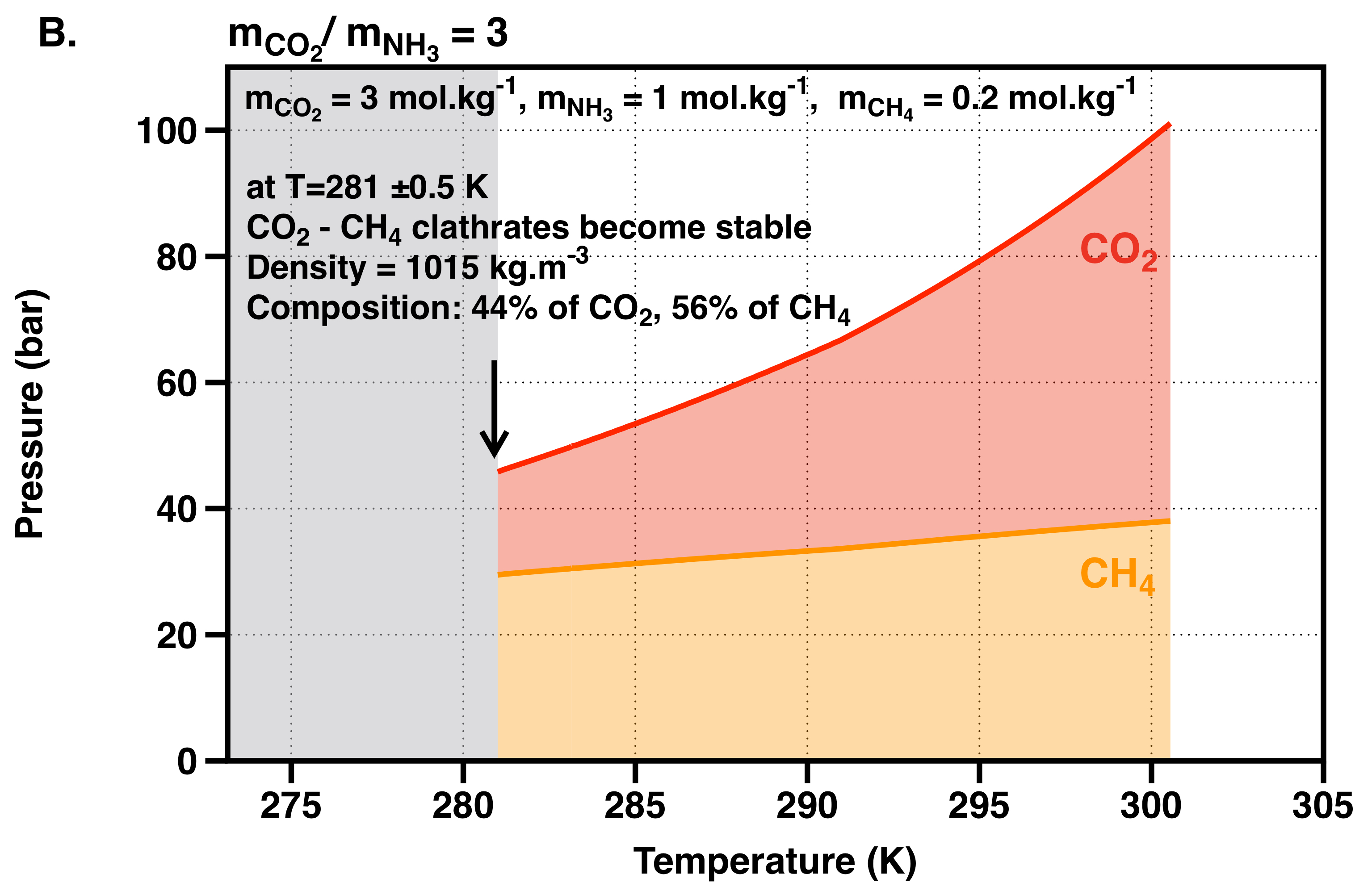}
\caption{\label{fig_clathrate1} Formation of CO$_2$-CH$_4$ mixed clathrates at Titan's surface for two $m_{CO_2}$/$m_{NH_3}$  molar fractions and a molality of CH$_4$ of 0.2 mol.kg$^{-1}$.  The grey colored area shows the range of temperature for which clathrate hydrates should form. In the stability domain of clathrate hydrates, no atmospheric pressure in equilibrium with clathrate phase is predicted. The red line represent the total atmospheric pressure, while the red and orange areas represent the contribution of P$_{CO_2}$ and P$_{CH_4}$ to the tatal atmospheric pressure. The composition and density of clathrate are just computed on the right boundary of the stability domain when clathrate hydrates start to form. In (A), the formation of the crust below 275 K impedes further exchanges between the atmosphere and the liquid reservoir. In (B), the clathrates being denser than the liquid, further cooling and exchanges are possible down to the melting curve of water. Pressures are displayed up to 100 bar of total surface pressure. Ammonia and water compounds are not plotted because they exist only in negligible quantities in the gas phase (see Fig. \ref{fig_resPL3}).}
\end{figure}

\begin{figure}
\centering
\includegraphics[width=0.85\linewidth]{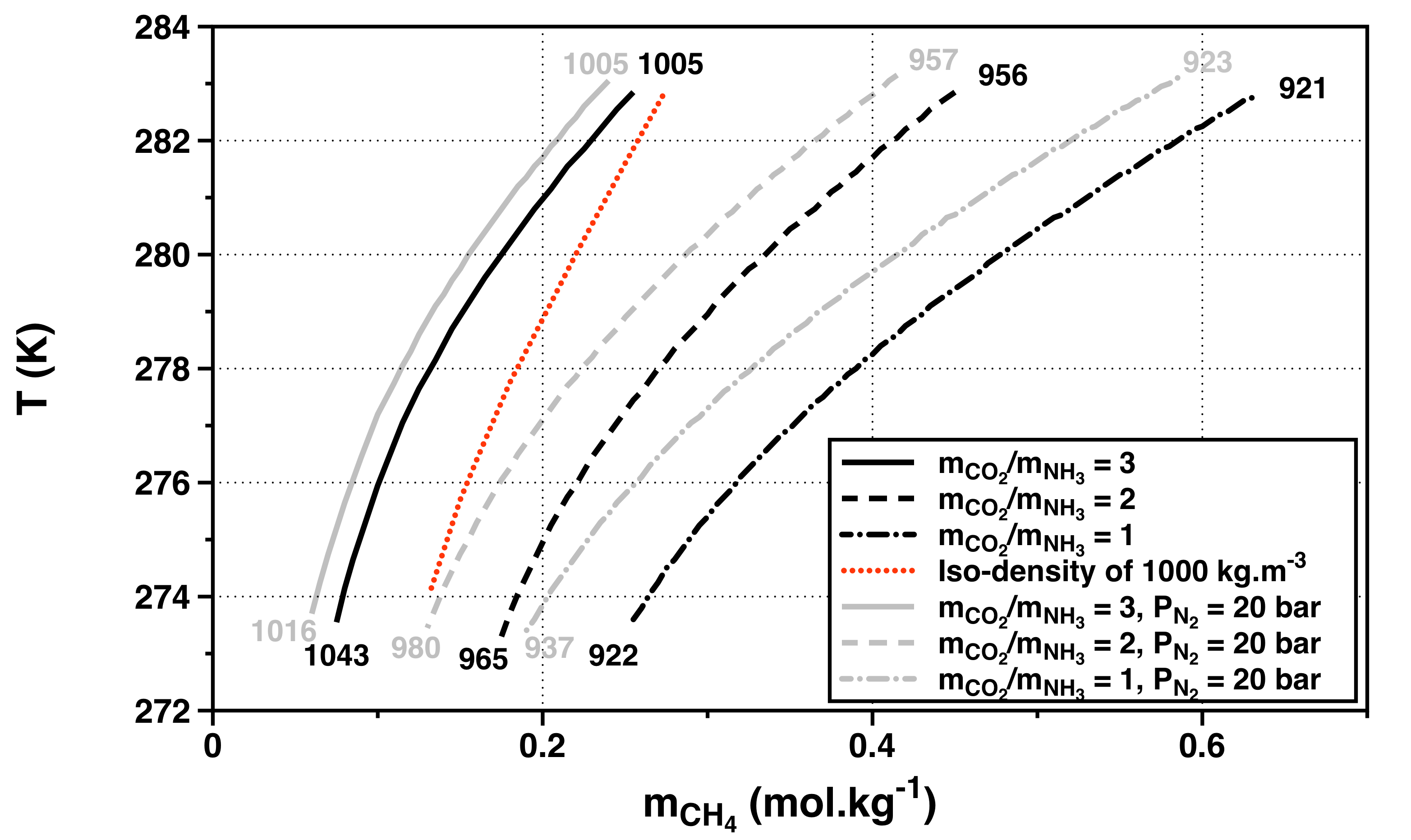}
\caption{\label{fig_clathrate2} Evolution of the temperature of clathrate-liquid-gas triple line, as a function of CH$_4$ mean molality, for three $m_{CO_2}$ /$m_{NH_3}$ ratios. Clathrate hydrate phase become stable at higher CH$_4$ molalities/lower temperatures from each triple line. Numbers indicate the clathrate densities at both sides of the triple lines. Red dotted line is the 1000 kg.m$^{-3}$ iso-density line for the case without N$_2$ in the atmosphere. For discussion purposes, the shifts of the triple lines, if 20 bar of N$_2$ partial pressure were added to the atmosphere, are indicated in grey.}
\end{figure}

Indeed, if the amount of CO$_2$ is sufficiently high for ensuring that a significant part of the carbon dioxide is trapped in the clathrate phase ($m_{CO_2} > m_{NH_3}$), both composition and densities of clathrate hydrates are changed up to a point where they may become denser than the liquid reservoir at the surface. In Figure \ref{fig_clathrate1}, the composition of the atmosphere is indicated for two examples as a function of temperature. Ammonia and water compounds are not plotted because they exist only in negligible quantities in the gas phase. The trend is comparable in both cases, with the atmosphere being mostly composed of the two gases (CO$_2$ and CH$_4$) with similar abundances, and with the existence of a critical temperature below which clathrates hydrates are formed (see Fig. \ref{fig_clathrate2} for $m_{CH_4}$=0.2). In Figure \ref{fig_clathrate1}A, where the $m_{CO_2}$ /$m_{NH_3}$  ratio is fixed to 2, clathrates are less dense than water and then should form a crust at Titan's surface. This should once again strongly reduce the possible exchanges between the two reservoirs, but the primordial atmosphere is then significantly enriched in carbon dioxide. In the second case (Fig. \ref{fig_clathrate1}B), where the $m_{CO_2}$ /$m_{NH_3}$  ratio is fixed to 3, the CO$_2$ abundance in the atmosphere is much more important and clathrates are significantly enriched in CO$_2$.\\  
Consequences of the atmospheric enrichment of CO$_2$ relative to CH$_4$ on the clathrate hydrates densities is illustrated in Figure \ref{fig_clathrate2}, where the densities of clathrates are computed from Equation \ref{eq_rho}. It turns out that in this second case the clathrates are denser than the liquid, and should fall down so that further exchanges between the liquid and the atmosphere should continue down to the freezing temperature of water ice.\\

In Figure \ref{fig_clathrate2}, the transition domain between the two regimes is plotted with the red curve that displays the surface density of the liquid reservoir. For the range of parameters explored in this study, it is extremely clear that one key parameter is the bulk abundance of methane. If CH$_4$ molality is above 0.3 mol.kg$^{-1}$, it seems that the crust will form very rapidly whatever the amount of carbon dioxide is. Even the adding of some nitrogen cannot change this conclusion. Possible exchanges of the atmosphere with the liquid reservoir down to very cold temperatures can be achieved only if the amount of methane is very low (for molalities lower than 0.1 mol.kg$^{-1}$). In between these two values, and as discussed just above, it is the amount of carbon dioxide with respect to ammonia that controls the density of the clathrate hydrates that form at low temperatures.\\

\section{Implication for Titan's primitive atmosphere and the formation of Titan's early crust }
\label{sec_implications}
Whatever the surface temperature is (above 273 K), there is no high pressure ices for $x_{melt}<$0.32. For the internal structures without a HP ice layer, the surface pressure depends on the volatile abundances of Titan's building blocks only, and do not depend on the $x_{melt}$ fraction. The maximal thickness of the ice VI layer is 290 km and is reached for  $x_{melt}$=0.5 and T=273.15 K, i.e. the case which is displayed in Figure \ref{fig_resPL3}. The value of $x_{melt}$ controls the amount of HP ice with respect to the liquid reservoir at a given temperature, and consequently the amount of volatiles in the system ocean-atmosphere since no volatiles are trapped in the HP ice layer. The formation of this layer leads to an increase of concentrations of dissolved species in the ocean-atmosphere system. Therefore, in this second case where the HP layer forms, the results do depend on the $x_{melt}$ fraction. However, this dependence implies negligible effects when compared to the uncertainties on the bulk composition in volatiles of the satellite and it does not affect our conclusions or the following discussion (see Fig. \ref{fig_resPL3}).\\

\subsection{The conversion of NH$_3$ into N$_2$ in the atmosphere}
In the previous section, we demonstrated that depending on the relative abundances of CO$_2$ and NH$_3$ in planetesimals that formed Titan, two scenarios may have occurred. For a global water ocean formed by impact heating, the repartition of the volatile compounds between the water ocean and the atmosphere are governed by the dissolution of gases in liquid water. We apply a thermodynamic model based on \citet{Darde2010} to model the dissolution of NH$_3$ and CO$_2$ in water and we show that if $m_{CO_2} < m_{NH_3}$, the partial pressure of CO$_2$ in the atmosphere remains low whatever the temperature is, and the partial pressure of NH$_3$ stays close to the saturation vapor pressure of the binary NH$_3$-H$_2$O system. On the contrary, for $m_{CO_2}> m_{NH_3}$, the partial pressure in the atmosphere of NH$_3$ is significantly reduced (see values displayed in Fig. \ref{fig_resPL3}) and the CO$_2$ is the main compound of Titan's primitive atmosphere. \\
\citet{Ishimaru2011} suggested that the conversion of NH$_3$ into N$_2$ by impacts would be favored for NH$_3$-rich atmospheres where the main atmospheric compound is CO$_2$ (CO$_2$:NH$_3$ ratios from 1:1 to 4:1 for a 1 bar atmosphere). Such atmospheric composition favors the NH$_3$ into N$_2$ conversion during impacts because of the low heat capacity of CO$_2$, allowing maintaining high temperatures over longer timescales than those obtained for impacts in reducing gases such as NH$_3$ or CH$_4$. However in our study, it has been shown that an atmospheric composition that is simultaneously rich in CO$_2$ and NH$_3$ is certainly not compatible with the presence of massive water ocean at the surface, as the excess of one chemical compound tends to favor the dissolution of the other one in water. \\

\subsection{Formation of methane-rich clathrate crust}
If the presence of clathrates allows the formation of a crust, it must be at a higher temperature than for the pure water ice crust. Thus, it implies an earlier separation of ocean and atmosphere and a rapid cessation of NH$_3$ into N$_2$ conversion in the atmosphere at low temperatures. In this study we explore the conditions of formation of CO$_2$-CH$_4$ clathrate hydrates, as well as the evolution of the primordial CH$_4$ at Titan's surface. Our results show that methane clathrates should form during cooling at the ocean-atmosphere interface, at low temperatures. The evolution of the formed CH$_4$-rich clathrates depends on the total amount of methane and also on the $m_{CO_2}/m_{NH_3}$ ratio. For very low CH$_4$ molalities ($<$0.1 mol.kg$^{-1}$), clathrate hydrates should sink as soon as they are formed. For higher values (above 0.3 mol.kg$^{-1}$), an upper crust should rapidly appear, thus strongly reducing exchanges between the atmosphere and the ocean. In between these values, it is the $m_{CO_2}/m_{NH_3}$ ratio that controls the system. For high $m_{CO_2}/m_{NH_3}$ ratios (roughly above 3), the clathrates possibly appearing at Titan's surface are CO$_2$-rich, denser than water, and consequently should sink in the ocean. On the contrary, for lower values of the $m_{CO_2}/m_{NH_3}$ ratio, CH$_4$-CO$_2$ clathrates with low CO$_2$ content are formed, thus resulting in the formation of an upper clathrate crust.\\ 

Primitive Titan's atmosphere may have contained up to 20 bar of N$_2$ if the NH$_3$ into N$_2$ conversion could occur at this epoch \citep{Atreya1978, Lunine1999, Adams2006, Atreya2010}. Adding N$_2$ to the atmosphere leads to two competing effects. On one hand, the increase of the total surface pressure favor the formation of clathrate since clathrate hydrates form when the total atmospheric pressure $P$ is higher than the pressure of formation of clathrates $P_d$  (eq.\ref{eq_mixclath}). On the other hand, the pressure of formation of the mixture of clathrates from eq. \ref{eq_mixclath} increase upon the addition of N$_2$ because the pressure of formation of N$_2$ pure clathrates ($>$ 159 bar, for liquid water+N$_2$) is an order of magnitude higher that the pressure of formation of pure CO$_2$ ($>$ 12 bar, for liquid water+CO$_2$) or pure CH$_4$ clathrates ($>$ 24 bar, for liquid water+CH$_4$). As an example, adding 20 bar of N$_2$ in to the total surface pressure increase the clathrate formation pressure $P_d$ of eq. \ref{eq_mixclath} by 8-10 bar (depending on the composition of the atmosphere prior the addition of N$_2$). \\

 The resulting effect for the formation of clathrates at the surface of Titan is shown in  Fig. \ref{fig_clathrate2}. The addition of 20 bar of N$_2$ tends to favor the clathrate formation, as they form at higher temperatures and higher methane concentration when compared to the case without N$_2$. This figure indicates that nitrogen should not affect the main results presented just above, but should increase the amount of clathrates hydrates formed in the primordial stages.

\section{Conclusion}
Considering a wide range of surface temperatures and global compositions of CO$_2$, NH$_3$ and CH$_4$ in Titan's building blocks, we compute the corresponding atmospheric compositions on early Titan in equilibrium with a global surface ocean.  We show that the $m_{CO_2}/m_{NH_3}$ ratio is crucial for the determination of the main atmospheric constituents, as NH$_3$ and CO$_2$ mutually increase their solubility in liquid water. For $m_{CO_2}/m_{NH_3}<$1, CO$_2$ would be massively dissolved in water, and its partial pressure won't exceed 1 bar for temperatures lower than 323K at Titan's surface. For $m_{CO_2}/m_{NH_3} >$1 the partial pressure can exceed more than 10 bars, making the partial pressure of CO$_2$ comparable or larger to that of CH$_4$. Assuming that Titan's initial volatile composition is comparable to that measured in Enceladus' plume \citep{Waite2017}, the $m_{CO_2}/m_{NH_3} $ ratio would be comprised between 0.2 and 2, and even larger values assuming cometary composition. Moreover, the observed abundances in Enceladus' plume may also reflect different solubilities of CO$_2$ and NH$_3$ in water. Due to an alkaline pH, most of the inorganic carbon might be dissolved as CO$_3^{2-}$ in Enceladus' inner ocean, leading for an initial CO$_2$ abundance much higher than the one currently observed in the plumes \citep{Zolotov2007, Glein2015a}.  In this case, the range of possible  $m_{CO_2}/m_{NH_3}$ could be extended to higher values. The lower estimate would imply an atmosphere dominated by CH$_4$  with small abundances of CO$_2$, while an atmosphere with several tens of bars of CO$_2$ and CH$_4$ could be generated for the upper bound. \\
We then showed that the abundance of CH$_4$ relative to CO$_2$ plays a crucial role in the clathration processes and the formation of the primordial crust. If the CH$_4$ abundance is lower than 0.1 mol.kg$^{-1}$, then the partial pressure of CH$_4$ in Titan's atmosphere is low relative to the partial pressure of CO$_2$ and clathrates formed at Titan's surface are CO$_2$-rich and denser than the liquid water. If the concentration of CH$_4$ in Titan's building blocks is higher than 0.3 mol.kg$^{-1}$, then the clathrates formed at Titan's surface are CH$_4$-rich and less dense than water and constitute a crust at Titan's surface, ceasing the further exchanges between the ocean and the atmosphere at temperatures higher than the melting temperature of water. Between 0.1 and 0.3 mol.kg$^{-1}$ of CH$_4$ concentration, the density of clathrates depend on the $m_{CO_2}/m_{NH_3}$ ratio in Titan's building blocks: a buoyant clathrate-rich crust is formed only if the $m_{CO_2}/m_{NH_3}$ ratio is lower than $\sim$3. \\
More investigations using an atmospheric radiative and photochemical model coupled with the chemical interaction model between the ocean and the atmosphere are required to estimate the amount of NH$_3$ that could be converted to N$_2$ before the formation of a crust at Titan's surface and form the present-day N$_2$-rich atmosphere. Considering the volatile abundances in comets and in Enceladus' plume, which indicate a CO$_2$ abundance comparable and higher than that of NH$_3$, future studies would probably need to take in account a CO$_2$-rich atmosphere for the early Titan, in order to evaluate the consequences it might have for the chemical evolution of this water-rich environment. Moreover, the volatile abundances of CH$_4$ in comets and Enceladus plume also indicate us that the formation of an early methane-rich crust at Titan's surface, before the surface of the satellite reaches the freezing point of water, could be a possible scenario. Our conclusions remain unchanged if we add a significant amount of N$_2$ that could originate from the NH$_3$ into N$_2$ conversion occurring in Titan's early atmosphere. \\
 
\section{Acknowledgments}
 The authors would like to thank Chris McKay and an anonymous reviewer for their insightful comments and suggestions that helped improve the manuscript. The research leading to these results has received funding from the European Research Council under the European Community's Seventh Framework Program (FP7/2007-2013 Grant Agreement no. 259285).

\section*{References}

\bibliography{library}

\begin{thebibliography}{84}
\expandafter\ifx\csname natexlab\endcsname\relax\def\natexlab#1{#1}\fi
\providecommand{\url}[1]{\texttt{#1}}
\providecommand{\href}[2]{#2}
\providecommand{\path}[1]{#1}
\providecommand{\DOIprefix}{doi:}
\providecommand{\ArXivprefix}{arXiv:}
\providecommand{\URLprefix}{URL: }
\providecommand{\Pubmedprefix}{pmid:}
\providecommand{\doi}[1]{\href{http://dx.doi.org/#1}{\path{#1}}}
\providecommand{\Pubmed}[1]{\href{pmid:#1}{\path{#1}}}
\providecommand{\bibinfo}[2]{#2}
\ifx\xfnm\relax \def\xfnm[#1]{\unskip,\space#1}\fi
\bibitem[{Adams(2006)}]{Adams2006}
\bibinfo{author}{Adams, E.}, \bibinfo{year}{2006}.
\newblock \bibinfo{title}{{Titan's Thermal Structure and the Formation of a
  Nitrogen Atmosphere}}.
\newblock Ph.D. thesis. University of Michigan.
\bibitem[{Al~Ghafri et~al.(2014)Al~Ghafri, Forte, Maitland,
  Rodriguez-Henríquez and Trusler}]{AlGhafri2014}
\bibinfo{author}{Al~Ghafri, S.Z.S.}, \bibinfo{author}{Forte, E.},
  \bibinfo{author}{Maitland, G.C.}, \bibinfo{author}{Rodriguez-Henríquez,
  J.J.}, \bibinfo{author}{Trusler, J.P.M.}, \bibinfo{year}{2014}.
\newblock \bibinfo{title}{Experimental and modeling study of the phase behavior
  of (methane + co2 + water) mixtures}.
\newblock \bibinfo{journal}{The Journal of Physical Chemistry B}
  \bibinfo{volume}{118}, \bibinfo{pages}{14461--14478}.
\newblock \DOIprefix\doi{10.1021/jp509678g},
  \href{http://arxiv.org/abs/http://dx.doi.org/10.1021/jp509678g}{\tt
  arXiv:http://dx.doi.org/10.1021/jp509678g}. \bibinfo{note}{pMID: 25406634}.
\bibitem[{{Alibert, Y.} and {Mousis, O.}(2007)}]{Alibert2007}
\bibinfo{author}{{Alibert, Y.}}, \bibinfo{author}{{Mousis, O.}},
  \bibinfo{year}{2007}.
\newblock \bibinfo{title}{Formation of titan in saturn's subnebula: constraints
  from huygens probe measurements}.
\newblock \bibinfo{journal}{Astronomy and Astrophysics} \bibinfo{volume}{465},
  \bibinfo{pages}{1051--1060}.
\newblock \DOIprefix\doi{10.1051/0004-6361:20066402}.
\bibitem[{Anderson and Prausnitz(1978)}]{Anderson1978}
\bibinfo{author}{Anderson, T.F.}, \bibinfo{author}{Prausnitz, J.M.},
  \bibinfo{year}{1978}.
\newblock \bibinfo{title}{{Application of the UNIQUAC equation to calculation
  of multicomponent phase equilibria. 1. Vapor-liquid equilibria}}.
\newblock \bibinfo{journal}{Industrial \& Engineering Chemistry Process Design
  and Development} \bibinfo{volume}{17}, \bibinfo{pages}{552--561}.
\bibitem[{Atreya et~al.(2006)Atreya, Adams, Niemann, Demick-Montelara, Owen,
  Fulchignoni, Ferri and Wilson}]{Atreya2006}
\bibinfo{author}{Atreya, S.K.}, \bibinfo{author}{Adams, E.Y.},
  \bibinfo{author}{Niemann, H.B.}, \bibinfo{author}{Demick-Montelara, J.E.},
  \bibinfo{author}{Owen, T.C.}, \bibinfo{author}{Fulchignoni, M.},
  \bibinfo{author}{Ferri, F.}, \bibinfo{author}{Wilson, E.H.},
  \bibinfo{year}{2006}.
\newblock \bibinfo{title}{Titan's methane cycle}.
\newblock \bibinfo{journal}{Planetary and Space Science} \bibinfo{volume}{54},
  \bibinfo{pages}{1177 -- 1187}.
\newblock \DOIprefix\doi{https://doi.org/10.1016/j.pss.2006.05.028}.
  \bibinfo{note}{{Surfaces and Atmospheres of the Outer Planets, their
  Satellites and Ring Systems from Cassini-Huygens DataEuropean Geosciences
  Union General Assembly - Sessions PS1.5, PS3.02 and PS3.03}}.
\bibitem[{Atreya et~al.(1978)Atreya, Donahue and Kuhn}]{Atreya1978}
\bibinfo{author}{Atreya, S.K.}, \bibinfo{author}{Donahue, T.},
  \bibinfo{author}{Kuhn, W.}, \bibinfo{year}{1978}.
\newblock \bibinfo{title}{{Evolution of a nitrogen atmosphere on Titan}}.
\newblock \bibinfo{journal}{Science} \bibinfo{volume}{201},
  \bibinfo{pages}{611--613}.
\bibitem[{Atreya and Lorenz(2010)}]{Atreya2010}
\bibinfo{author}{Atreya, S.K.}, \bibinfo{author}{Lorenz, Ralph D.and~Waite,
  J.H.}, \bibinfo{year}{2010}.
\newblock \bibinfo{title}{Volatile Origin and Cycles: Nitrogen and Methane}.
  \bibinfo{publisher}{Springer Netherlands}. chapter~\bibinfo{chapter}{7}.
\newblock pp. \bibinfo{pages}{177--199}.
\newblock \DOIprefix\doi{10.1007/978-1-4020-9215-2_7}.
\bibitem[{Bamberger et~al.(2000)Bamberger, Sieder and Maurer}]{Bamberger2000}
\bibinfo{author}{Bamberger, a.}, \bibinfo{author}{Sieder, G.},
  \bibinfo{author}{Maurer, G.}, \bibinfo{year}{2000}.
\newblock \bibinfo{title}{{High-pressure (vapor + liquid) equilibrium in binary
  mixtures of (carbon dioxide + water or acetic acid) at temperatures from 313
  to 353 K}}.
\newblock \bibinfo{journal}{Journal of Supercritical Fluids}
  \bibinfo{volume}{17}, \bibinfo{pages}{97--110}.
\newblock \DOIprefix\doi{10.1016/S0896-8446(99)00054-6}.
\bibitem[{Bockel{\'{e}}e-Morvan and Crovisier(2004)}]{Bockelee-Morvan2004}
\bibinfo{author}{Bockel{\'{e}}e-Morvan, D.}, \bibinfo{author}{Crovisier, J.},
  \bibinfo{year}{2004}.
\newblock \bibinfo{title}{{The composition of cometary volatiles}}, in:
  \bibinfo{booktitle}{Comets II}. \bibinfo{publisher}{University of Arizona
  Press}, pp. \bibinfo{pages}{391--424}.
\bibitem[{Choukroun and Grasset(2007)}]{Choukroun2007}
\bibinfo{author}{Choukroun, M.}, \bibinfo{author}{Grasset, O.},
  \bibinfo{year}{2007}.
\newblock \bibinfo{title}{Thermodynamic model for water and high-pressure ices
  up to 2.2 gpa and down to the metastable domain}.
\newblock \bibinfo{journal}{The Journal of chemical physics}
  \bibinfo{volume}{127}, \bibinfo{pages}{124506}.
\bibitem[{Choukroun and Grasset(2010)}]{Choukroun2013}
\bibinfo{author}{Choukroun, M.}, \bibinfo{author}{Grasset, O.},
  \bibinfo{year}{2010}.
\newblock \bibinfo{title}{{Thermodynamic data and modeling of the water and
  ammonia-water phase diagrams up to 2.2 GPa for planetary geophysics.}}
\newblock \bibinfo{journal}{The Journal of chemical physics}
  \bibinfo{volume}{133}, \bibinfo{pages}{144502}.
\newblock \DOIprefix\doi{10.1063/1.3487520}.
\bibitem[{Choukroun et~al.(2010)Choukroun, Grasset, Tobie and
  Sotin}]{Choukroun2010}
\bibinfo{author}{Choukroun, M.}, \bibinfo{author}{Grasset, O.},
  \bibinfo{author}{Tobie, G.}, \bibinfo{author}{Sotin, C.},
  \bibinfo{year}{2010}.
\newblock \bibinfo{title}{{Stability of methane clathrate hydrates under
  pressure: Influence on outgassing processes of methane on Titan}}.
\newblock \bibinfo{journal}{Icarus} \bibinfo{volume}{205},
  \bibinfo{pages}{581--593}.
\newblock \DOIprefix\doi{10.1016/j.icarus.2009.08.011}.
\bibitem[{Choukroun et~al.(2013)Choukroun, Kieffer, Lu and
  Tobie}]{Choukroun2013a}
\bibinfo{author}{Choukroun, M.}, \bibinfo{author}{Kieffer, S.W.},
  \bibinfo{author}{Lu, X.}, \bibinfo{author}{Tobie, G.}, \bibinfo{year}{2013}.
\newblock \bibinfo{title}{{Clathrate Hydrates : Implications for Exchange
  Processes in the outer Solar System}}, in: \bibinfo{editor}{Gudipati, M.S.},
  \bibinfo{editor}{Castillo-Rogez, J.} (Eds.), \bibinfo{booktitle}{The Science
  of Solar System Ices}.
\bibitem[{Cochran et~al.(2015)Cochran, Levasseur-Regourd, Cordiner, Hadamcik,
  Lasue, Gicquel, Schleicher, Charnley, Mumma, Paganini et~al.}]{Cochran2015}
\bibinfo{author}{Cochran, A.L.}, \bibinfo{author}{Levasseur-Regourd, A.C.},
  \bibinfo{author}{Cordiner, M.}, \bibinfo{author}{Hadamcik, E.},
  \bibinfo{author}{Lasue, J.}, \bibinfo{author}{Gicquel, A.},
  \bibinfo{author}{Schleicher, D.G.}, \bibinfo{author}{Charnley, S.B.},
  \bibinfo{author}{Mumma, M.J.}, \bibinfo{author}{Paganini, L.}, et~al.,
  \bibinfo{year}{2015}.
\newblock \bibinfo{title}{The composition of comets}.
\newblock \bibinfo{journal}{Space Science Reviews} \bibinfo{volume}{197},
  \bibinfo{pages}{9--46}.
\bibitem[{Cramer(1984)}]{Cramer1984}
\bibinfo{author}{Cramer, S.D.}, \bibinfo{year}{1984}.
\newblock \bibinfo{title}{{Solubility of methane in brines from 0 to 300C}}.
\newblock \bibinfo{journal}{Industrial {\&} Engineering Chemistry Process
  Design and Development} \bibinfo{volume}{23}, \bibinfo{pages}{533--538}.
\newblock \DOIprefix\doi{10.1021/i200026a021}.
\bibitem[{Darde et~al.(2012)Darde, Maribo-Mogensen, van Well, Stenby and
  Thomsen}]{Darde2012}
\bibinfo{author}{Darde, V.}, \bibinfo{author}{Maribo-Mogensen, B.},
  \bibinfo{author}{van Well, W.J.M.}, \bibinfo{author}{Stenby, E.H.},
  \bibinfo{author}{Thomsen, K.}, \bibinfo{year}{2012}.
\newblock \bibinfo{title}{{Process simulation of CO2 capture with aqueous
  ammonia using the Extended UNIQUAC model}}.
\newblock \bibinfo{journal}{International Journal of Greenhouse Gas Control}
  \bibinfo{volume}{10}, \bibinfo{pages}{74--87}.
\newblock \DOIprefix\doi{10.1016/j.ijggc.2012.05.017}.
\bibitem[{Darde et~al.(2010)Darde, van Well, Stenby and Thomsen}]{Darde2010}
\bibinfo{author}{Darde, V.}, \bibinfo{author}{van Well, W.J.M.},
  \bibinfo{author}{Stenby, E.H.}, \bibinfo{author}{Thomsen, K.},
  \bibinfo{year}{2010}.
\newblock \bibinfo{title}{{Modeling of carbon dioxide absorption by aqueous
  ammonia solutions using the Extended UNIQUAC model}}.
\newblock \bibinfo{journal}{Industrial {\&} Engineering Chemistry Research}
  \bibinfo{volume}{49}, \bibinfo{pages}{12663--12674}.
\newblock \DOIprefix\doi{10.1021/ie1009519}.
\bibitem[{Dhima et~al.(1999)Dhima, Hemptinne and Jose}]{Dhima1999}
\bibinfo{author}{Dhima, A.}, \bibinfo{author}{Hemptinne, J.D.},
  \bibinfo{author}{Jose, J.}, \bibinfo{year}{1999}.
\newblock \bibinfo{title}{{Solubility of hydrocarbons and CO2 mixtures in water
  under high pressure}}.
\newblock \bibinfo{journal}{Industrial \& engineering chemistry research}
  \bibinfo{volume}{38}, \bibinfo{pages}{3144--3161}.
\bibitem[{Edwards et~al.(1978)Edwards, Maurer, Newman and
  Prausnitz}]{Edwards1978}
\bibinfo{author}{Edwards, T.J.}, \bibinfo{author}{Maurer, G.},
  \bibinfo{author}{Newman, J.}, \bibinfo{author}{Prausnitz, J.M.},
  \bibinfo{year}{1978}.
\newblock \bibinfo{title}{Vapor-liquid equilibria in multicomponent aqueous
  solutions of volatile weak electrolytes}.
\newblock \bibinfo{journal}{AIChe Journal} \bibinfo{volume}{24},
  \bibinfo{pages}{966--976}.
\bibitem[{Edwards et~al.(1975)Edwards, Newman and Prausnitz}]{Edwards1975}
\bibinfo{author}{Edwards, T.J.}, \bibinfo{author}{Newman, J.},
  \bibinfo{author}{Prausnitz, J.M.}, \bibinfo{year}{1975}.
\newblock \bibinfo{title}{Thermodynamics of aqueous solutions containing
  volatile weak electrolytes}.
\newblock \bibinfo{journal}{AIChe Journal} \bibinfo{volume}{21},
  \bibinfo{pages}{248--259}.
\bibitem[{Englezos(1993)}]{Englezos1993}
\bibinfo{author}{Englezos, P.}, \bibinfo{year}{1993}.
\newblock \bibinfo{title}{{The clathrate hydrates}}.
\newblock \bibinfo{journal}{Industrial {\&} Engineering Chemistry Research}
  \bibinfo{volume}{32}, \bibinfo{pages}{1251--1274}.
\bibitem[{Estrada et~al.(2009)Estrada, Mosqueira, Lissauer, D’Angelo and
  Cruikshank}]{Estrada2009}
\bibinfo{author}{Estrada, P.R.}, \bibinfo{author}{Mosqueira, I.},
  \bibinfo{author}{Lissauer, J.}, \bibinfo{author}{D’Angelo, G.},
  \bibinfo{author}{Cruikshank, D.}, \bibinfo{year}{2009}.
\newblock \bibinfo{title}{Formation of jupiter and conditions for accretion of
  the galilean satellites}.
\newblock \bibinfo{journal}{Europa, edited by RT Pappalardo, WB McKinnon, and
  K. Khurana, University of Arizona Press, Tucson} , \bibinfo{pages}{27--58}.
\bibitem[{Gasem et~al.(2001)Gasem, Gao, Pan and Robinson}]{Gasem2001}
\bibinfo{author}{Gasem, K.A.M.}, \bibinfo{author}{Gao, W.},
  \bibinfo{author}{Pan, Z.}, \bibinfo{author}{Robinson, R.L.},
  \bibinfo{year}{2001}.
\newblock \bibinfo{title}{{A modified temperature dependence for the
  Peng-Robinson equation of state}}.
\newblock \bibinfo{journal}{Fluid Phase Equilibria} \bibinfo{volume}{181},
  \bibinfo{pages}{113--125}.
\newblock \DOIprefix\doi{10.1016/S0378-3812(01)00488-5}.
\bibitem[{Glein et~al.(2008)Glein, Zolotov and Shock}]{Glein2008}
\bibinfo{author}{Glein, C.}, \bibinfo{author}{Zolotov, M.},
  \bibinfo{author}{Shock, E.}, \bibinfo{year}{2008}.
\newblock \bibinfo{title}{{The oxidation state of hydrothermal systems on early
  Enceladus}}.
\newblock \bibinfo{journal}{Icarus} \bibinfo{volume}{197},
  \bibinfo{pages}{157--163}.
\newblock \DOIprefix\doi{10.1016/j.icarus.2008.03.021}.
\bibitem[{Glein(2015)}]{Glein2015a}
\bibinfo{author}{Glein, C.R.}, \bibinfo{year}{2015}.
\newblock \bibinfo{title}{{Noble gases, nitrogen, and methane from the deep
  interior to the atmosphere of Titan}}.
\newblock \bibinfo{journal}{Icarus} \bibinfo{volume}{250},
  \bibinfo{pages}{570--586}.
\newblock \DOIprefix\doi{10.1016/j.icarus.2015.01.001}.
\bibitem[{Glein(2017)}]{Glein2017}
\bibinfo{author}{Glein, C.R.}, \bibinfo{year}{2017}.
\newblock \bibinfo{title}{A whiff of nebular gas in titan's atmosphere –
  potential implications for the conditions and timing of titan's formation}.
\newblock \bibinfo{journal}{Icarus} \bibinfo{volume}{293}, \bibinfo{pages}{231
  -- 242}.
\newblock \DOIprefix\doi{https://doi.org/10.1016/j.icarus.2017.02.026}.
\bibitem[{Glein et~al.(2009)Glein, Desch and Shock}]{Glein2009}
\bibinfo{author}{Glein, C.R.}, \bibinfo{author}{Desch, S.J.},
  \bibinfo{author}{Shock, E.L.}, \bibinfo{year}{2009}.
\newblock \bibinfo{title}{{The absence of endogenic methane on Titan and its
  implications for the origin of atmospheric nitrogen}}.
\newblock \bibinfo{journal}{Icarus} \bibinfo{volume}{204},
  \bibinfo{pages}{637--644}.
\newblock \DOIprefix\doi{10.1016/j.icarus.2009.06.020}.
\bibitem[{G{\"{o}}ppert and Maurer(1988)}]{Goppert1988}
\bibinfo{author}{G{\"{o}}ppert, U.}, \bibinfo{author}{Maurer, G.},
  \bibinfo{year}{1988}.
\newblock \bibinfo{title}{{Vapor-liquid equilibria in aqueous solutions of
  ammonia and carbon dioxide at temperatures between 333 and 393 K and
  pressures up to 7 MPa}}.
\newblock \bibinfo{journal}{Fluid Phase Equilibria} \bibinfo{volume}{41},
  \bibinfo{pages}{153--185}.
\bibitem[{Griffith et~al.(2013)Griffith, Mitchell, Lavvas and
  Tobie}]{Griffith2013}
\bibinfo{author}{Griffith, C.}, \bibinfo{author}{Mitchell, J.},
  \bibinfo{author}{Lavvas, P.}, \bibinfo{author}{Tobie, G.},
  \bibinfo{year}{2013}.
\newblock \bibinfo{title}{{Titan's Evolving Climate}}, in:
  \bibinfo{booktitle}{Comparative Climatology of Terrestrial Planets}, pp.
  \bibinfo{pages}{1--27}.
\bibitem[{Harms-Watzenberg(1995)}]{HarmsWatzenberg1995}
\bibinfo{author}{Harms-Watzenberg, F.}, \bibinfo{year}{1995}.
\newblock \bibinfo{title}{Measurement and correlation of the thermodynamic
  properties of water-ammonia mixtures}.
\newblock \bibinfo{journal}{VDI Progress Reports} .
\bibitem[{Hersant et~al.(2008)Hersant, Gautier, Tobie and Lunine}]{Hersant2008}
\bibinfo{author}{Hersant, F.}, \bibinfo{author}{Gautier, D.},
  \bibinfo{author}{Tobie, G.}, \bibinfo{author}{Lunine, J.I.},
  \bibinfo{year}{2008}.
\newblock \bibinfo{title}{{Interpretation of the carbon abundance in Saturn
  measured by Cassini}}.
\newblock \bibinfo{journal}{Planetary and Space Science} \bibinfo{volume}{56},
  \bibinfo{pages}{1103--1111}.
\newblock \DOIprefix\doi{10.1016/j.pss.2008.02.007}.
\bibitem[{H{\"{o}}rst et~al.(2008)H{\"{o}}rst, Vuitton and Yelle}]{Horst2008}
\bibinfo{author}{H{\"{o}}rst, S.M.}, \bibinfo{author}{Vuitton, V.},
  \bibinfo{author}{Yelle, R.V.}, \bibinfo{year}{2008}.
\newblock \bibinfo{title}{{Origin of oxygen species in Titan's atmosphere}}.
\newblock \bibinfo{journal}{Journal of Geophysical Research}
  \bibinfo{volume}{113}.
\newblock \DOIprefix\doi{10.1029/2008JE003135}.
\bibitem[{Ishimaru et~al.(2011)Ishimaru, Sekine, Matsui and
  Mousis}]{Ishimaru2011}
\bibinfo{author}{Ishimaru, R.}, \bibinfo{author}{Sekine, Y.},
  \bibinfo{author}{Matsui, T.}, \bibinfo{author}{Mousis, O.},
  \bibinfo{year}{2011}.
\newblock \bibinfo{title}{{Oxidizing proto-atmosphere on Titan: Constraint from
  N2 formation by impact shock}}.
\newblock \bibinfo{journal}{The Astrophysical Journal} \bibinfo{volume}{741},
  \bibinfo{pages}{L10}.
\newblock \DOIprefix\doi{10.1088/2041-8205/741/1/L10}.
\bibitem[{Jilvero et~al.(2015)Jilvero, Jens, Normann, Andersson, Halstensen,
  Eimer and Johnsson}]{Jilvero2015}
\bibinfo{author}{Jilvero, H.}, \bibinfo{author}{Jens, K.J.},
  \bibinfo{author}{Normann, F.}, \bibinfo{author}{Andersson, K.},
  \bibinfo{author}{Halstensen, M.}, \bibinfo{author}{Eimer, D.},
  \bibinfo{author}{Johnsson, F.}, \bibinfo{year}{2015}.
\newblock \bibinfo{title}{{Equilibrium measurements of the NH3--CO2--H2O system
  -- measurement and evaluation of vapor liquid equilibrium data at low
  temperatures}}.
\newblock \bibinfo{journal}{Fluid Phase Equilibria} \bibinfo{volume}{385},
  \bibinfo{pages}{237--247}.
\newblock \DOIprefix\doi{10.1016/j.fluid.2014.11.006}.
\bibitem[{Krasnopolsky(2010)}]{Krasnopolsky2010}
\bibinfo{author}{Krasnopolsky, V.A.}, \bibinfo{year}{2010}.
\newblock \bibinfo{title}{{The photochemical model of Titan's atmosphere and
  ionosphere: A version without hydrodynamic escape}}.
\newblock \bibinfo{journal}{Planetary and Space Science} \bibinfo{volume}{58},
  \bibinfo{pages}{1507--1515}.
\newblock \DOIprefix\doi{10.1016/j.pss.2010.07.010}.
\bibitem[{Kuramoto and Matsui(1994)}]{Kuramoto1994}
\bibinfo{author}{Kuramoto, K.}, \bibinfo{author}{Matsui, T.},
  \bibinfo{year}{1994}.
\newblock \bibinfo{title}{Formation of a hot proto-atmosphere on the accreting
  giant icy satellite: Implications for the origin and evolution of titan,
  ganymede, and callisto}.
\newblock \bibinfo{journal}{Journal of Geophysical Research: Planets}
  \bibinfo{volume}{99}, \bibinfo{pages}{21183--21200}.
\bibitem[{Kurz et~al.(1995)Kurz, Rumpf and Maurer}]{Kurz1995}
\bibinfo{author}{Kurz, F.}, \bibinfo{author}{Rumpf, B.},
  \bibinfo{author}{Maurer, G.}, \bibinfo{year}{1995}.
\newblock \bibinfo{title}{{Vapor-liquid-solid equilibria in the system
  NH3-CO2-H2O from around 310 to 470 K: New experimental data and modeling}}.
\newblock \bibinfo{journal}{Fluid Phase Equilibria} \bibinfo{volume}{104},
  \bibinfo{pages}{261--275}.
\bibitem[{Lavvas et~al.(2011)Lavvas, Galand, Yelle, Heays, Lewis, Lewis and
  Coates}]{Lavvas2011}
\bibinfo{author}{Lavvas, P.}, \bibinfo{author}{Galand, M.},
  \bibinfo{author}{Yelle, R.}, \bibinfo{author}{Heays, A.},
  \bibinfo{author}{Lewis, B.}, \bibinfo{author}{Lewis, G.},
  \bibinfo{author}{Coates, A.}, \bibinfo{year}{2011}.
\newblock \bibinfo{title}{{Energy deposition and primary chemical products in
  Titan’s upper atmosphere}}.
\newblock \bibinfo{journal}{Icarus} \bibinfo{volume}{213},
  \bibinfo{pages}{233--251}.
\bibitem[{{Linstrom, P. J. Mallard}(2016)}]{NIST_webbook}
\bibinfo{author}{{Linstrom, P. J. Mallard}, W.G.}, \bibinfo{year}{2016}.
\newblock \bibinfo{title}{{NIST WebBook of Chemistry, Standart refernce
  database num. 69}}.
\newblock \bibinfo{address}{National Institute of Standards and Technology,
  Gaithersburg MD, 20899}.
\newblock \DOIprefix\doi{doi:10.18434/T4D303}.
\bibitem[{Lipenkov and Istomin(2001)}]{Lipenkov2001}
\bibinfo{author}{Lipenkov, V.Y.}, \bibinfo{author}{Istomin, V.},
  \bibinfo{year}{2001}.
\newblock \bibinfo{title}{On the stability of air clathrate-hydrate crystals in
  subglacial lake vostok, antarctica}.
\newblock \bibinfo{journal}{Materialy Glyatsiol. Issled} \bibinfo{volume}{91},
  \bibinfo{pages}{2001}.
\bibitem[{Lunine and Lorenz(2009)}]{Lunine2009}
\bibinfo{author}{Lunine, J.I.}, \bibinfo{author}{Lorenz, R.D.},
  \bibinfo{year}{2009}.
\newblock \bibinfo{title}{{Rivers, Lakes, Dunes, and Rain: Crustal Processes in
  Titan's Methane Cycle}}.
\newblock \bibinfo{journal}{Annual Review of Earth and Planetary Sciences}
  \bibinfo{volume}{37}, \bibinfo{pages}{299--320}.
\newblock \DOIprefix\doi{10.1146/annurev.earth.031208.100142}.
\bibitem[{Lunine and Stevenson(1987)}]{Lunine1987}
\bibinfo{author}{Lunine, J.I.}, \bibinfo{author}{Stevenson, D.J.},
  \bibinfo{year}{1987}.
\newblock \bibinfo{title}{{Clathrate and ammonia hydrates at high pressure:
  Application to the origin of methane on Titan}}.
\newblock \bibinfo{journal}{Icarus} \bibinfo{volume}{70},
  \bibinfo{pages}{61--77}.
\newblock \DOIprefix\doi{10.1016/0019-1035(87)90075-3}.
\bibitem[{Lunine et~al.(1999)Lunine, Yung and Lorenz}]{Lunine1999}
\bibinfo{author}{Lunine, J.I.}, \bibinfo{author}{Yung, Y.L.},
  \bibinfo{author}{Lorenz, R.D.}, \bibinfo{year}{1999}.
\newblock \bibinfo{title}{{On the volatile inventory of Titan from isotopic
  abundances in nitrogen and methane.}}
\newblock \bibinfo{journal}{Planetary and space science} \bibinfo{volume}{47},
  \bibinfo{pages}{1291--303}.
\bibitem[{Mandt et~al.(2014)Mandt, Mousis, Lunine and Gautier}]{Mandt2014}
\bibinfo{author}{Mandt, K.E.}, \bibinfo{author}{Mousis, O.},
  \bibinfo{author}{Lunine, J.}, \bibinfo{author}{Gautier, D.},
  \bibinfo{year}{2014}.
\newblock \bibinfo{title}{{Protosolar ammonia as the unique source of Titan's
  nitrigen}}.
\newblock \bibinfo{journal}{The Astrophysical Journal} \bibinfo{volume}{788},
  \bibinfo{pages}{L24}.
\newblock \DOIprefix\doi{10.1088/2041-8205/788/2/L24}.
\bibitem[{Mandt et~al.(2012)Mandt, Waite, Teolis, Magee, Bell, Westlake, Nixon,
  Mousis and Lunine}]{Mandt2012}
\bibinfo{author}{Mandt, K.E.}, \bibinfo{author}{Waite, J.H.},
  \bibinfo{author}{Teolis, B.}, \bibinfo{author}{Magee, B.a.},
  \bibinfo{author}{Bell, J.}, \bibinfo{author}{Westlake, J.H.},
  \bibinfo{author}{Nixon, C.a.}, \bibinfo{author}{Mousis, O.},
  \bibinfo{author}{Lunine, J.I.}, \bibinfo{year}{2012}.
\newblock \bibinfo{title}{{The 12C/13C ratio on Titan from Cassini INMS
  measurements and implications for the evolution of methane}}.
\newblock \bibinfo{journal}{The Astrophysical Journal} \bibinfo{volume}{749}.
\newblock \DOIprefix\doi{10.1088/0004-637X/749/2/160}.
\bibitem[{Marounina et~al.(2015)Marounina, Tobie, Carpy, Monteux, Charnay and
  Grasset}]{Marounina2015}
\bibinfo{author}{Marounina, N.}, \bibinfo{author}{Tobie, G.},
  \bibinfo{author}{Carpy, S.}, \bibinfo{author}{Monteux, J.},
  \bibinfo{author}{Charnay, B.}, \bibinfo{author}{Grasset, O.},
  \bibinfo{year}{2015}.
\newblock \bibinfo{title}{{Evolution of Titan's atmosphere during the Late
  Heavy Bombardment}}.
\newblock \bibinfo{journal}{Icarus} \bibinfo{volume}{257},
  \bibinfo{pages}{324--335}.
\newblock \DOIprefix\doi{10.1016/j.icarus.2015.05.011}.
\bibitem[{Matson et~al.(2007)Matson, Castillo, Lunine and Johnson}]{Matson2007}
\bibinfo{author}{Matson, D.L.}, \bibinfo{author}{Castillo, J.C.},
  \bibinfo{author}{Lunine, J.}, \bibinfo{author}{Johnson, T.V.},
  \bibinfo{year}{2007}.
\newblock \bibinfo{title}{{Enceladus' plume: Compositional evidence for a hot
  interior}}.
\newblock \bibinfo{journal}{Icarus} \bibinfo{volume}{187},
  \bibinfo{pages}{569--573}.
\newblock \DOIprefix\doi{10.1016/j.icarus.2006.10.016}.
\bibitem[{McCord et~al.(2008)McCord, Hayne, Combe, Hansen, Barnes, Rodriguez,
  {Le Mou{\'{e}}lic}, Baines, Buratti, Sotin, Nicholson, Jaumann, Nelson and
  {the Cassini VIMS Team}}]{McCord2008}
\bibinfo{author}{McCord, T.B.}, \bibinfo{author}{Hayne, P.},
  \bibinfo{author}{Combe, J.P.}, \bibinfo{author}{Hansen, G.B.},
  \bibinfo{author}{Barnes, J.W.}, \bibinfo{author}{Rodriguez, S.},
  \bibinfo{author}{{Le Mou{\'{e}}lic}, S.}, \bibinfo{author}{Baines, E.K.H.},
  \bibinfo{author}{Buratti, B.J.}, \bibinfo{author}{Sotin, C.},
  \bibinfo{author}{Nicholson, P.}, \bibinfo{author}{Jaumann, R.},
  \bibinfo{author}{Nelson, R.}, \bibinfo{author}{{the Cassini VIMS Team}},
  \bibinfo{year}{2008}.
\newblock \bibinfo{title}{{Titan's surface: Search for spectral diversity and
  composition using the Cassini VIMS investigation}}. volume
  \bibinfo{volume}{194}.
\newblock \DOIprefix\doi{10.1016/j.icarus.2007.08.039}.
\bibitem[{McKay et~al.(2003)McKay, Hand, Doran, Andersen and
  Priscu}]{McKay2003}
\bibinfo{author}{McKay, C.P.}, \bibinfo{author}{Hand, K.P.},
  \bibinfo{author}{Doran, P.T.}, \bibinfo{author}{Andersen, D.T.},
  \bibinfo{author}{Priscu, J.C.}, \bibinfo{year}{2003}.
\newblock \bibinfo{title}{{Clathrate formation and the fate of noble and
  biologically useful gases in Lake Vostok, Antarctica}}.
\newblock \bibinfo{journal}{Geophysical Research Letters} \bibinfo{volume}{30},
  \bibinfo{pages}{1702}.
\newblock \DOIprefix\doi{10.1029/2003GL017490}.
\bibitem[{McKay et~al.(1988)McKay, Scattergood, Pollack, Borucki and {Van
  Ghyseghem}}]{McKay1988}
\bibinfo{author}{McKay, C.P.}, \bibinfo{author}{Scattergood, T.},
  \bibinfo{author}{Pollack, J.B.}, \bibinfo{author}{Borucki, W.J.},
  \bibinfo{author}{{Van Ghyseghem}, H.T.}, \bibinfo{year}{1988}.
\newblock \bibinfo{title}{{High-temperature shock formation of N2 and organics
  on primordial Titan}}.
\newblock \bibinfo{journal}{Nature} \bibinfo{volume}{332},
  \bibinfo{pages}{520--522}.
\bibitem[{Miller(1974)}]{Miller1974}
\bibinfo{author}{Miller, S.L.}, \bibinfo{year}{1974}.
\newblock \bibinfo{title}{The Nature and Occurrence of Clathrate Hydrates}.
  \bibinfo{publisher}{Springer US}, \bibinfo{address}{Boston, MA}.
\newblock pp. \bibinfo{pages}{151--177}.
\newblock \DOIprefix\doi{10.1007/978-1-4684-2757-8_10}.
\bibitem[{Monteux et~al.(2014)Monteux, Tobie, Choblet and {Le
  Feuvre}}]{Monteux2014}
\bibinfo{author}{Monteux, J.}, \bibinfo{author}{Tobie, G.},
  \bibinfo{author}{Choblet, G.}, \bibinfo{author}{{Le Feuvre}, M.},
  \bibinfo{year}{2014}.
\newblock \bibinfo{title}{{Can large icy moons accrete undifferentiated?}}
\newblock \bibinfo{journal}{Icarus} \bibinfo{volume}{237},
  \bibinfo{pages}{377--387}.
\newblock \DOIprefix\doi{10.1016/j.icarus.2014.04.041}.
\bibitem[{Mousis et~al.(2009)Mousis, Lunine, Thomas, Pasek, Marb{\oe}uf,
  Alibert, Ballenegger, Cordier, Ellinger, Pauzat et~al.}]{Mousis2009b}
\bibinfo{author}{Mousis, O.}, \bibinfo{author}{Lunine, J.I.},
  \bibinfo{author}{Thomas, C.}, \bibinfo{author}{Pasek, M.},
  \bibinfo{author}{Marb{\oe}uf, U.}, \bibinfo{author}{Alibert, Y.},
  \bibinfo{author}{Ballenegger, V.}, \bibinfo{author}{Cordier, D.},
  \bibinfo{author}{Ellinger, Y.}, \bibinfo{author}{Pauzat, F.}, et~al.,
  \bibinfo{year}{2009}.
\newblock \bibinfo{title}{Clathration of volatiles in the solar nebula and
  implications for the origin of titan's atmosphere}.
\newblock \bibinfo{journal}{The Astrophysical Journal} \bibinfo{volume}{691},
  \bibinfo{pages}{1780}.
\bibitem[{M{\"{u}}ller et~al.(1988)M{\"{u}}ller, Bender and
  Maurer}]{Muller1988}
\bibinfo{author}{M{\"{u}}ller, G.}, \bibinfo{author}{Bender, E.},
  \bibinfo{author}{Maurer, G.}, \bibinfo{year}{1988}.
\newblock \bibinfo{title}{{Das Dampf-Fl{\"{u}}ssigkeitsgleichgenwicht des
  tern{\"{a}}ren Systems Ammoniak-Kohlendioxid-Wasser bei hohen Wassergehalten
  im Bereich zwischen 373 un 474 Kelvin}}.
\newblock \bibinfo{journal}{Berichte der Bunsengesellschaft f{\"{u}}r
  Physikalische Chemie} \bibinfo{volume}{92}, \bibinfo{pages}{148--160}.
\bibitem[{Mumma and Charnley(2011)}]{Mumma2011}
\bibinfo{author}{Mumma, M.J.}, \bibinfo{author}{Charnley, S.B.},
  \bibinfo{year}{2011}.
\newblock \bibinfo{title}{{The Chemical Composition of Comets -- Emerging
  Taxonomies and Natal Heritage}}.
\newblock \bibinfo{journal}{Annual Review of Astronomy and Astrophysics}
  \bibinfo{volume}{49}, \bibinfo{pages}{471--524}.
\newblock \DOIprefix\doi{10.1146/annurev-astro-081309-130811}.
\bibitem[{Nicolaisen et~al.(1993)Nicolaisen, Rasmussen and
  S{\o}rensen}]{Nicolaisen1993}
\bibinfo{author}{Nicolaisen, H.}, \bibinfo{author}{Rasmussen, P.},
  \bibinfo{author}{S{\o}rensen, J.M.}, \bibinfo{year}{1993}.
\newblock \bibinfo{title}{{Correlation and prediction of mineral solubilities
  in the reciprocal salt system (Na+, K+)(Cl-, SO2-4)-H2O at 0-100°C}}.
\newblock \bibinfo{journal}{Chemical Engineering Science} \bibinfo{volume}{48},
  \bibinfo{pages}{3149--3158}.
\newblock \DOIprefix\doi{10.1016/0009-2509(93)80201-Z}.
\bibitem[{Niemann et~al.(2005)Niemann, Atreya, Bauer, Carignan, Demick, Frost,
  Gautier, Haberman, Harpold, Hunten, Israel, Lunine, Kasprzak, Owen,
  Paulkovich, Raulin, Raaen and Way}]{Niemann2005}
\bibinfo{author}{Niemann, H.B.}, \bibinfo{author}{Atreya, S.K.},
  \bibinfo{author}{Bauer, S.J.}, \bibinfo{author}{Carignan, G.R.},
  \bibinfo{author}{Demick, J.E.}, \bibinfo{author}{Frost, R.L.},
  \bibinfo{author}{Gautier, D.}, \bibinfo{author}{Haberman, J.A.},
  \bibinfo{author}{Harpold, D.N.}, \bibinfo{author}{Hunten, D.M.},
  \bibinfo{author}{Israel, G.}, \bibinfo{author}{Lunine, J.I.},
  \bibinfo{author}{Kasprzak, W.T.}, \bibinfo{author}{Owen, T.C.},
  \bibinfo{author}{Paulkovich, M.}, \bibinfo{author}{Raulin, F.},
  \bibinfo{author}{Raaen, E.}, \bibinfo{author}{Way, S.H.},
  \bibinfo{year}{2005}.
\newblock \bibinfo{title}{{The abundances of constituents of Titan's atmosphere
  from the GCMS instrument on the Huygens probe.}}
\newblock \bibinfo{journal}{Nature} \bibinfo{volume}{438},
  \bibinfo{pages}{779--784}.
\newblock \DOIprefix\doi{10.1038/nature04122}.
\bibitem[{Niemann et~al.(2010)Niemann, Atreya, Demick, Gautier, Haberman,
  Harpold, Kasprzak, Lunine, Owen and Raulin}]{Niemann2010}
\bibinfo{author}{Niemann, H.B.}, \bibinfo{author}{Atreya, S.K.},
  \bibinfo{author}{Demick, J.E.}, \bibinfo{author}{Gautier, D.},
  \bibinfo{author}{Haberman, J.a.}, \bibinfo{author}{Harpold, D.N.},
  \bibinfo{author}{Kasprzak, W.T.}, \bibinfo{author}{Lunine, J.I.},
  \bibinfo{author}{Owen, T.C.}, \bibinfo{author}{Raulin, F.},
  \bibinfo{year}{2010}.
\newblock \bibinfo{title}{{Composition of Titan's lower atmosphere and simple
  surface volatiles as measured by the Cassini-Huygens probe gas chromatograph
  mass spectrometer experiment}}.
\newblock \bibinfo{journal}{Journal of Geophysical Research}
  \bibinfo{volume}{115}, \bibinfo{pages}{E12006}.
\newblock \DOIprefix\doi{10.1029/2010JE003659}.
\bibitem[{Owen et~al.(2006)Owen, Niemann, Atreya and Zolotov}]{Owen2006}
\bibinfo{author}{Owen, T.C.}, \bibinfo{author}{Niemann, H.},
  \bibinfo{author}{Atreya, S.}, \bibinfo{author}{Zolotov, M.Y.},
  \bibinfo{year}{2006}.
\newblock \bibinfo{title}{{Between heaven and Earth: the exploration of
  Titan}}.
\newblock \bibinfo{journal}{Faraday Discussions} \bibinfo{volume}{133},
  \bibinfo{pages}{387--391}.
\newblock \DOIprefix\doi{10.1039/b517174a}.
\bibitem[{Pazuki et~al.(2006)Pazuki, Pahlevanzadeh and Ahooei}]{Pazuki2006}
\bibinfo{author}{Pazuki, G.}, \bibinfo{author}{Pahlevanzadeh, H.},
  \bibinfo{author}{Ahooei, A.M.}, \bibinfo{year}{2006}.
\newblock \bibinfo{title}{Prediction of phase behavior of co 2--nh 3--h 2 o
  system by using the uniquac-non random factor (nrf) model}.
\newblock \bibinfo{journal}{Fluid phase equilibria} \bibinfo{volume}{242},
  \bibinfo{pages}{57--64}.
\bibitem[{Peng and Robinson(1976)}]{Peng1976}
\bibinfo{author}{Peng, D.Y.}, \bibinfo{author}{Robinson, D.B.},
  \bibinfo{year}{1976}.
\newblock \bibinfo{title}{{A New Two-Constant Equation of State}}.
\newblock \bibinfo{journal}{Industrial {\&} Engineering Chemistry Fundamentals}
  \bibinfo{volume}{15}, \bibinfo{pages}{59--64}.
\newblock \DOIprefix\doi{10.1021/i160057a011}.
\bibitem[{Prausnitz(1963)}]{Prausnitz1963}
\bibinfo{author}{Prausnitz, J.}, \bibinfo{year}{1963}.
\newblock \bibinfo{title}{Thermodynamic representation of high-pressure
  vapour-liquid equilibria}.
\newblock \bibinfo{journal}{Chemical Engineering Science} \bibinfo{volume}{18},
  \bibinfo{pages}{613--630}.
\bibitem[{Prausnitz et~al.(1998)Prausnitz, Lichtenthaler and
  de~Azevedo}]{Prausnitz1998}
\bibinfo{author}{Prausnitz, J.}, \bibinfo{author}{Lichtenthaler, R.},
  \bibinfo{author}{de~Azevedo, E.}, \bibinfo{year}{1998}.
\newblock \bibinfo{title}{Molecular Thermodynamics of Fluid-Phase Equilibria}.
\newblock Prentice Hall international series in the physical and chemical
  engineering sciences, \bibinfo{publisher}{Prentice-Hall PTR}.
\bibitem[{Prinn and Fegley~Jr(1981)}]{Prinn1981}
\bibinfo{author}{Prinn, R.G.}, \bibinfo{author}{Fegley~Jr, B.},
  \bibinfo{year}{1981}.
\newblock \bibinfo{title}{Kinetic inhibition of co and n2 reduction in
  circumplanetary nebulae-implications for satellite composition}.
\newblock \bibinfo{journal}{The Astrophysical Journal} \bibinfo{volume}{249},
  \bibinfo{pages}{308--317}.
\bibitem[{Rousselot et~al.(2013)Rousselot, Pirali, Jehin, Vervloet,
  Hutsem{\'{e}}kers, Manfroid, Cordier, Martin-Drumel, Gruet, Arpigny, Decock
  and Mousis}]{Rousselot2014}
\bibinfo{author}{Rousselot, P.}, \bibinfo{author}{Pirali, O.},
  \bibinfo{author}{Jehin, E.}, \bibinfo{author}{Vervloet, M.},
  \bibinfo{author}{Hutsem{\'{e}}kers, D.}, \bibinfo{author}{Manfroid, J.},
  \bibinfo{author}{Cordier, D.}, \bibinfo{author}{Martin-Drumel, M.A.},
  \bibinfo{author}{Gruet, S.}, \bibinfo{author}{Arpigny, C.},
  \bibinfo{author}{Decock, A.}, \bibinfo{author}{Mousis, O.},
  \bibinfo{year}{2013}.
\newblock \bibinfo{title}{{Toward a Unique Nitrogen Isotopic Ratio in Cometary
  Ices}}.
\newblock \bibinfo{journal}{The Astrophysical Journal Letters}
  \bibinfo{volume}{780}, \bibinfo{pages}{L17}.
\newblock \DOIprefix\doi{10.1088/2041-8205/780/2/L17}.
\bibitem[{Rumpf and Maurer(1993a)}]{Rumpf1993co2}
\bibinfo{author}{Rumpf, B.}, \bibinfo{author}{Maurer, G.},
  \bibinfo{year}{1993}a.
\newblock \bibinfo{title}{{An experimental and theoretical investigation on the
  solubility of carbon dioxide in aqueous solutions of strong electrolytes}}.
\newblock \bibinfo{journal}{Berichte der Bunsengesellschaft fuer physikalische
  chemie} \bibinfo{volume}{97}, \bibinfo{pages}{85--97}.
\bibitem[{Rumpf and Maurer(1993b)}]{Rumpf1993nh3}
\bibinfo{author}{Rumpf, B.}, \bibinfo{author}{Maurer, G.},
  \bibinfo{year}{1993}b.
\newblock \bibinfo{title}{Solubility of ammonia in aqueous solutions of sodium
  sulfate and ammonium sulfate at temperatures from 333.15 k to 433.15 k and
  pressures up to 3 mpa}.
\newblock \bibinfo{journal}{Industrial \& engineering chemistry research}
  \bibinfo{volume}{32}, \bibinfo{pages}{1780--1789}.
\bibitem[{Sander et~al.(1986)Sander, Rasmussen and Fredenslund}]{Sander1986}
\bibinfo{author}{Sander, B.}, \bibinfo{author}{Rasmussen, P.},
  \bibinfo{author}{Fredenslund, A.}, \bibinfo{year}{1986}.
\newblock \bibinfo{title}{{Calculation of vapour-liquid equilibria in nitric
  acid-water-nitrate salt systems using an extended UNIQUAC equation}}.
\newblock \bibinfo{journal}{Chemical engineering science} \bibinfo{volume}{41},
  \bibinfo{pages}{1185--1195}.
\bibitem[{Sekine et~al.(2011)Sekine, Genda, Sugita, Kadono and
  Matsui}]{Sekine2011}
\bibinfo{author}{Sekine, Y.}, \bibinfo{author}{Genda, H.},
  \bibinfo{author}{Sugita, S.}, \bibinfo{author}{Kadono, T.},
  \bibinfo{author}{Matsui, T.}, \bibinfo{year}{2011}.
\newblock \bibinfo{title}{{Replacement and late formation of atmospheric N2 on
  undifferentiated Titan by impacts}}.
\newblock \bibinfo{journal}{Nature Geoscience} \bibinfo{volume}{4},
  \bibinfo{pages}{359--362}.
\newblock \DOIprefix\doi{10.1038/NGEO1147}.
\bibitem[{Shinnaka et~al.(2014)Shinnaka, Kawakita, Kobayashi, Nagashima and
  Boice}]{Shinnaka2014}
\bibinfo{author}{Shinnaka, Y.}, \bibinfo{author}{Kawakita, H.},
  \bibinfo{author}{Kobayashi, H.}, \bibinfo{author}{Nagashima, M.},
  \bibinfo{author}{Boice, D.C.}, \bibinfo{year}{2014}.
\newblock \bibinfo{title}{{14NH2/15NH2 Ratio in Comet C/2012 S1 (Ison) Observed
  During Its Outburst in 2013 November}}.
\newblock \bibinfo{journal}{The Astrophysical Journal} \bibinfo{volume}{782},
  \bibinfo{pages}{L16}.
\newblock \DOIprefix\doi{10.1088/2041-8205/782/2/L16}.
\bibitem[{Sloan and Koh(2008)}]{Sloan2008}
\bibinfo{author}{Sloan, D.}, \bibinfo{author}{Koh, C.A.}, \bibinfo{year}{2008}.
\newblock \bibinfo{title}{Clathrate Hydrates of Natural Gases, 3rd edition}.
\newblock \bibinfo{publisher}{CRC Press}.
\bibitem[{Smolen et~al.(1991)Smolen, Manley and Poling}]{Smolen1991}
\bibinfo{author}{Smolen, T.M.}, \bibinfo{author}{Manley, D.B.},
  \bibinfo{author}{Poling, B.E.}, \bibinfo{year}{1991}.
\newblock \bibinfo{title}{{Vapor-Liquid Equilibrium Data for the Ammonia-water
  System and Its Description with a Modified Cubic Equation of State}}.
\newblock \bibinfo{journal}{Journal of Chemical {\&} Engineering Data}
  \bibinfo{volume}{36}, \bibinfo{pages}{202--208}.
\newblock \DOIprefix\doi{10.1021/je00002a017}.
\bibitem[{Spycher et~al.(2003)Spycher, Pruess and Ennis-King}]{Spycher2003}
\bibinfo{author}{Spycher, N.}, \bibinfo{author}{Pruess, K.},
  \bibinfo{author}{Ennis-King, J.}, \bibinfo{year}{2003}.
\newblock \bibinfo{title}{{CO2-H2O mixtures in the geological sequestration of
  CO2. I. Assessment and calculation of mutual solubilities from 12 to 100°C
  and up to 600 bar}}.
\newblock \bibinfo{journal}{Geochimica et Cosmochimica Acta}
  \bibinfo{volume}{67}, \bibinfo{pages}{3015--3031}.
\newblock \DOIprefix\doi{10.1016/S0016-7037(03)00273-4}.
\bibitem[{Thomsen(2005)}]{Thomsen2005}
\bibinfo{author}{Thomsen, K.}, \bibinfo{year}{2005}.
\newblock \bibinfo{title}{{Modeling electrolyte solutions with the extended
  universal quasichemical (UNIQUAC) model}}.
\newblock \bibinfo{journal}{Pure and Applied Chemistry} \bibinfo{volume}{77},
  \bibinfo{pages}{531--542}.
\newblock \DOIprefix\doi{10.1351/pac200577030531}.
\bibitem[{Thomsen and Rasmussen(1999)}]{Thomsen1999}
\bibinfo{author}{Thomsen, K.}, \bibinfo{author}{Rasmussen, P.},
  \bibinfo{year}{1999}.
\newblock \bibinfo{title}{{Modeling of vapor–-liquid-–solid equilibrium in
  gas–-aqueous electrolyte systems}}.
\newblock \bibinfo{journal}{Chemical Engineering Science} \bibinfo{volume}{54},
  \bibinfo{pages}{1787--1802}.
\newblock \DOIprefix\doi{10.1016/S0009-2509(99)00019-6}.
\bibitem[{Tobie et~al.(2012)Tobie, Gautier and Hersant}]{Tobie2012}
\bibinfo{author}{Tobie, G.}, \bibinfo{author}{Gautier, D.},
  \bibinfo{author}{Hersant, F.}, \bibinfo{year}{2012}.
\newblock \bibinfo{title}{{Titan'S Bulk Composition Constrained By
  Cassini-Huygens : Implication for Internal Outgassing}}.
\newblock \bibinfo{journal}{The Astrophysical Journal} \bibinfo{volume}{752},
  \bibinfo{pages}{125}.
\newblock \DOIprefix\doi{10.1088/0004-637X/752/2/125}.
\bibitem[{Tobie et~al.(2013)Tobie, Lunine, Monteux, Mousis and
  Nimmo}]{Tobie2014}
\bibinfo{author}{Tobie, G.}, \bibinfo{author}{Lunine, J.},
  \bibinfo{author}{Monteux, J.}, \bibinfo{author}{Mousis, O.},
  \bibinfo{author}{Nimmo, F.}, \bibinfo{year}{2013}.
\newblock \bibinfo{title}{The origin and evolution of titan}.
\newblock \bibinfo{journal}{Titan: Interior, Surface, Atmosphere and Space
  Environment} , \bibinfo{pages}{24--55}.
\bibitem[{Tobie et~al.(2006)Tobie, Lunine and Sotin}]{Tobie2006}
\bibinfo{author}{Tobie, G.}, \bibinfo{author}{Lunine, J.I.},
  \bibinfo{author}{Sotin, C.}, \bibinfo{year}{2006}.
\newblock \bibinfo{title}{{Episodic outgassing as the origin of atmospheric
  methane on Titan.}}
\newblock \bibinfo{journal}{Nature} \bibinfo{volume}{440},
  \bibinfo{pages}{61--64}.
\newblock \DOIprefix\doi{10.1038/nature04497}.
\bibitem[{Van~Krevelen et~al.(1949)Van~Krevelen, Hoftijzer and
  Huntjens}]{VanKrevelen1949}
\bibinfo{author}{Van~Krevelen, D.W.}, \bibinfo{author}{Hoftijzer, P.J.},
  \bibinfo{author}{Huntjens, F.J.}, \bibinfo{year}{1949}.
\newblock \bibinfo{title}{{Composition and vapour pressures of aqueous
  solutions of ammmonia, carbon dioxide and hydrogen sulphide}}.
\newblock \bibinfo{journal}{Recueil des Travaux Chimiques des Pays-Bas}
  \bibinfo{volume}{68}, \bibinfo{pages}{191--216}.
\newblock \DOIprefix\doi{10.1002/recl.19490680213}.
\bibitem[{Verbrugge(1973)}]{Verbrugge1979}
\bibinfo{author}{Verbrugge, P.}, \bibinfo{year}{1973}.
\newblock \bibinfo{title}{{Vapour-liquid equilibria of the ammonia-carbon
  dioxide-water system}}.
\newblock Ph.D. thesis.
\bibitem[{Waite et~al.(2017)Waite, Glein, Perryman, Teolis, Magee, Miller,
  Grimes, Perry, Miller, Bouquet, Lunine, Brockwell and Bolton}]{Waite2017}
\bibinfo{author}{Waite, J.H.}, \bibinfo{author}{Glein, C.R.},
  \bibinfo{author}{Perryman, R.S.}, \bibinfo{author}{Teolis, B.D.},
  \bibinfo{author}{Magee, B.A.}, \bibinfo{author}{Miller, G.},
  \bibinfo{author}{Grimes, J.}, \bibinfo{author}{Perry, M.E.},
  \bibinfo{author}{Miller, K.E.}, \bibinfo{author}{Bouquet, A.},
  \bibinfo{author}{Lunine, J.I.}, \bibinfo{author}{Brockwell, T.},
  \bibinfo{author}{Bolton, S.J.}, \bibinfo{year}{2017}.
\newblock \bibinfo{title}{Cassini finds molecular hydrogen in the enceladus
  plume: Evidence for hydrothermal processes}.
\newblock \bibinfo{journal}{Science} \bibinfo{volume}{356},
  \bibinfo{pages}{155--159}.
\newblock \DOIprefix\doi{10.1126/science.aai8703}.
\bibitem[{Wilson and Atreya(2004)}]{Wilson2004}
\bibinfo{author}{Wilson, E.H.}, \bibinfo{author}{Atreya, S.K.},
  \bibinfo{year}{2004}.
\newblock \bibinfo{title}{{Current state of modeling the photochemistry of
  Titan's mutually dependent atmosphere and ionosphere}}.
\newblock \bibinfo{journal}{Journal of Geophysical Research}
  \bibinfo{volume}{109}, \bibinfo{pages}{E06002}.
\newblock \DOIprefix\doi{10.1029/2003JE002181}.
\bibitem[{Yelle et~al.(2008)Yelle, Cui and M{\"u}ller-Wodarg}]{Yelle2008}
\bibinfo{author}{Yelle, R.}, \bibinfo{author}{Cui, J.},
  \bibinfo{author}{M{\"u}ller-Wodarg, I.}, \bibinfo{year}{2008}.
\newblock \bibinfo{title}{{Methane escape from Titan's atmosphere}}.
\newblock \bibinfo{journal}{Journal of Geophysical Research: Planets}
  \bibinfo{volume}{113}.
\bibitem[{Zolotov(2007)}]{Zolotov2007}
\bibinfo{author}{Zolotov, M.Y.}, \bibinfo{year}{2007}.
\newblock \bibinfo{title}{{An oceanic composition on early and today's
  Enceladus}}.
\newblock \bibinfo{journal}{Geophysical Research Letters} \bibinfo{volume}{34},
  \bibinfo{pages}{1--5}.
\newblock \DOIprefix\doi{10.1029/2007GL031234}.

\end{thebibliography}

\end{document}